\documentclass[%
reprint,
nofootinbib,
 amsmath,amssymb,
 aps,
prd,
floatfix,
]{revtex4-2}

\usepackage{graphicx}
\usepackage{dcolumn}
\usepackage{bm}
\usepackage{booktabs} 
\usepackage{hyperref}
\usepackage{xcolor}
\definecolor{medium-blue}{rgb}{0,0,1}
\usepackage[font=small,labelfont=bf,
   justification=raggedright,
   format=plain]{caption}
\usepackage{float}
\usepackage{subcaption}
\usepackage{pifont}

\newcommand{\bigoh}[1]{\mathcal{O}(#1)}
\begin{document}

\preprint{APS/123-QED}

\title{Time-domain reconstruction of signals and glitches in gravitational wave data with deep learning}

\author{Tom Dooney$^{1,2,3}$}
\author{Harsh Narola$^{2,3}$}
\author{Stefano Bromuri$^{1}$}
\author{R. Lyana Curier$^{1}$}
\author{Chris Van Den Broeck$^{2,3}$}
\author{Sarah Caudill$^{4}$}
\author{Daniel Stanley Tan$^{1}$}

\affiliation{$^1$Faculty of Science, Open Universiteit, Valkenburgerweg 177, 6419 AT Heerlen, The Netherlands}
\affiliation{$^2$Institute for Gravitational and Subatomic Physics (GRASP), Utrecht University, Princetonplein 1, 3584 CC, Utrecht, The Netherlands}
\affiliation{$^3$Nikhef, Science Park 105,
1098 XG, Amsterdam, The Netherlands.}
\affiliation{$^4$Department of Physics, University of Massachusetts, Dartmouth, MA 02747, USA}


\begin{abstract}
Gravitational wave (GW) detectors, such as LIGO, Virgo, and KAGRA, detect faint signals from distant astrophysical events. 
However, their high sensitivity also makes them susceptible to background noise, which can obscure these signals. 
This noise often includes transient artifacts called `glitches', that can mimic genuine astrophysical signals or mask their true characteristics.
In this study, we present \textit{DeepExtractor}, a deep learning framework that is designed to reconstruct signals and glitches with power exceeding interferometer noise, regardless of their source. 
We design \textit{DeepExtractor} to model the inherent noise distribution of GW detectors, following conventional assumptions that the noise is Gaussian and stationary over short time scales.
It operates by predicting and subtracting the noise component of the data, retaining only the clean reconstruction of signal or glitch.
We focus on applications related to glitches and validate \textit{DeepExtractor’s} effectiveness through three experiments: (1) reconstructing simulated glitches injected into simulated detector noise, (2) comparing its performance with the state-of-the-art \textit{BayesWave} algorithm, and (3) analyzing real data from the \textit{Gravity Spy} dataset to demonstrate effective glitch subtraction from LIGO strain data.
We further demonstrate its potential by reconstructing three real GW events from LIGO’s third observing run, without being trained on GW waveforms.
Our proposed model achieves a median mismatch of only $0.9\%$ for simulated glitches, outperforming several deep learning baselines. Additionally, \textit{DeepExtractor} surpasses \textit{BayesWave} in glitch recovery, offering a dramatic computational speedup by reconstructing one glitch sample in approximately $0.1$ seconds on a CPU, compared to \textit{BayesWave’s} processing time of approximately one hour per glitch.
\end{abstract}

\maketitle

\section{\label{sec:Introduction}Introduction}
The first detection of gravitational waves (GWs) from a binary black hole (BBH) merger in 2015 ushered in a new era of astrophysics \cite{first_GW}. Since then, Advanced LIGO \cite{LIGO_paper} and Advanced Virgo \cite{VIRGOpaper} detectors have made confident detections of $\sim90$ compact binary coalescence (CBC) GW events \cite{GWTC1, GWTC_2_catalog, GWTC_3_catalog, 1-OGC, 2-OGC, 3-OGC, New_search_1, New_search_2, New_search_3}, accumulated over three observing runs (O1, O2, O3).
With the fourth observing (O4) run currently underway, hundreds of more detections are expected from the enhanced sensitivity of GW detectors \cite{Detchar_nextgen_glitch}.

GW data analysis procedures, such as those regarding the estimation of GW source parameters \cite{PE_1, PE_2, PE_3}, typically assume the detector noise to be stationary and Gaussian \cite{detchar_transient}.
This assumption breaks down in the presence of `glitches'; transient noise artifacts that introduce non-Gaussian and non-stationary features in the data \cite{ligo_detchar_o3, virgo_detchar_o3, kagra_detchar}.
Glitches are unmodeled terrestrial noise events stemming from environmental factors (e.g.,
earthquakes, wind, anthropogenic noise) or instrumental issues (e.g., control systems, electronic components \cite{Soni_2021_}). 
However, the origins of many glitches remain unknown \cite{detchar_transient}.

Model-free algorithms, \textit{Omicron} \cite{Omicron} and \textit{Gravity Spy} \cite{Zevin_2017}, are utilized for glitch detection and characterization, respectively.
These tools visualize and identify glitches by analyzing excess power in Q-scans (spectrograms) \cite{Q_scan}, which are time-frequency representations.
Glitches exhibit a broad range of time-frequency morphologies, which \textit{Gravity Spy} classifies into over 20 distinct categories.

Certain glitches can resemble astrophysical signals (eg. \textit{blip} glitches have similar time-frequency morphology to high-mass BBH signals), increasing false positives ~\cite{abbott2018effects, KAGRA:2022dwb,Steltner:2023cfk, KAGRA:2021kbb,Steltner:2021qjy,gw_subtract_2,driggers2019improving, blackburn2008lsc, abbott2016characterization} and hindering subsequent analyses. 
However, when glitches overlap with a GW signal they can significantly bias parameter estimation \cite{pankow2018mitigation, PE_glitch_1, PE_glitch_2} and must be carefully modeled and removed before the GW signal can be used to make any measurements.
This is a commonplace occurrence, with $\sim25$ out of the $\sim90$ confident detections so far requiring some form of glitch mitigation \cite{GWTC_3_catalog, pankow2018mitigation, BNS_1, BNS_2, BNS_BBH_1, BNS_BBH_2}. 

Ongoing upgrades to current detector systems and the introduction of third-generation detectors, such as the Einstein Telescope (ET) \cite{ET_paper} and Cosmic Explorer (CE) \cite{Cosmic_expl}, will significantly increase the detection rates from all sources of GWs. 
For instance, it is estimated that the ET alone will detect on the order of $\bigoh{10^4} \, \text{y}^{-1}$ BBH mergers and binary neutron star (BNS) mergers \cite{Iacovelli_2022, Kalogera_BNS}.
This could lead to over 400 GW events per day. 
If glitch rates remain similar (approximately 1 per minute during the third observing run \cite{GWTC_3_catalog, GWTC_2_catalog}), higher detection rates will naturally exacerbate issues relating to glitches \cite{Detchar_nextgen_glitch}.  
Such challenges demand fast reconstruction algorithms that can accurately separate events from the underlying detector background, agnostic of their sources.

In this study, we present \textit{DeepExtractor}, a power spectral density (PSD)-informed deep learning framework designed to accurately reconstruct power excesses above a Gaussian detector background in the time domain. 
By simultaneously analyzing magnitude and phase spectrograms from the short-time Fourier transform (STFT), \textit{DeepExtractor} isolates the underlying noise from signal and glitch events using a U-Net architecture \cite{Unet_original}. 
The signal or glitch is then reconstructed by subtracting the predicted noise, following an additive model.
Assuming additive interactions and linear approximations of signal and glitch components using time-domain wavelets, we simulate five waveform classes. 
By learning to isolate background noise from these classes, \textit{DeepExtractor} is capable of interpolating to arbitrary signals and glitches.

\textit{DeepExtractor} demonstrates superior accuracy in reconstructing simulated glitches and generalizes well to unseen distributions, outperforming several deep learning models.
We also present a comparison of glitch reconstructions by \textit{DeepExtractor} to those obtained using the state-of-the-art \textit{BayesWave} \cite{BayesWave}, an advanced algorithm used in GW data analysis to model and subtract glitches from the strain data.
We further demonstrate \textit{DeepExtractor's} effectiveness in reconstructing real \textit{Gravity Spy} \cite{Zevin_2017} glitches from LIGO’s third observing run, verified through Q-scans and time series plots of the detector data before and after glitch subtraction.
This marks the first comprehensive application of deep learning for such tasks in GW data analysis.
Finally, we use \textit{DeepExtractor} to reconstruct three real GW events detected by LIGO during O3, showcasing its ability to generalize to GW signals without being explicitly trained on these waveforms.

This paper is structured as follows; in Section \ref{sec:Related_Work}, we discuss the state-of-the-art in glitch modelling and mitigation in LIGO and Virgo and state-of-the-art machine learning research for signal and glitch reconstruction.
In Section \ref{sec:Methods}, we discuss our method, which includes an overview of our framework, our model architecture, the generation of training data, and the experiments implemented to show \textit{DeepExtractor's} efficacy.
In Section \ref{sec:Results}, we present the results of the experiments and Section \ref{sec:Conclusions} discusses the conclusions of our research.

\section{\label{sec:Related_Work}Related Work}

\subsection{\label{sec:GW_glitches}GW glitch mitigation}
A variety of methods have been developed to model and subtract glitches from the strain data that leverage machine learning \cite{ML_glitch_2} and parametric \cite{glitch_parametric, ArchEnemy_scattered_light_removal} models. 
However, up to the end of O3, \textit{BayesWave} \cite{BayesWave, Bayes_Glitch_1, pankow2018mitigation, Bayes_Glitch_5} and \textit{gwsubtract} \cite{gwsubtract_1, gw_subtract_2, Davis_O3_glitches} have been the only algorithms used for modeling and subtracting glitches in LIGO-Virgo-KAGRA data analyses \cite{Davis_O3_glitches}.
While \textit{gwsubtract} relies on strongly correlated auxiliary witness sensors to linearly subtract glitches from the strain data, \textit{BayesWave} models glitches using the detector strain only.

\textit{BayesWave} models the strain $h(t)$ in each detector as a linear combination: $h(t) = n(t) + s(t) + g(t)$,
where $n(t)$ represents Gaussian noise, $s(t)$ an astrophysical signal, and $g(t)$ a glitch. 
Non-Gaussian components, including signals and glitches, are represented as sums of sine-Gaussian (Morlet-Gabor) wavelets (see Eq. $4$ of \cite{BayesWave}).

\textit{BayesWave} performs Bayesian inference to estimate the wavelet parameters and the total number of wavelets of a glitch using a Reversible-Jump Markov Chain Monte Carlo (RJMCMC) sampler, which allows for the flexibility of models with variable dimensions.

For the glitch model, wavelet parameters are independent across detectors. For the signal model, a single set of wavelets is projected onto detectors based on extrinsic parameters such as the source sky location and polarization. The PSD is modeled with a combination of a cubic spline and Lorentzians, using the method from \cite{Bayes_noise}.

Three model assumptions guide the reconstruction of transient features in the data:
\begin{enumerate}
    \item Gaussian noise with an astrophysical signal.
    \item Gaussian noise with an instrumental glitch.
    \item Gaussian noise with both a signal and a glitch.
\end{enumerate}

Glitch subtraction uses either the glitch-only model or the signal+glitch model. The glitch-only model applies when the glitch is temporally or spectrally distinct from the signal. 
The signal+glitch model, used for significant overlaps, relies on data from multiple detectors to separate coherent signal power from incoherent glitch power, ensuring the astrophysical signal is preserved.

\textit{BayesWave} outputs a posterior distribution of the model parameters, derived from the RJMCMC. A random sample from this posterior distribution is used to reconstruct the time domain glitch, which is subtracted from the strain data to mitigate the glitch \cite{Bayes_Glitch_5}.

\begin{figure*}[t]
\centering
\includegraphics[width = \textwidth]{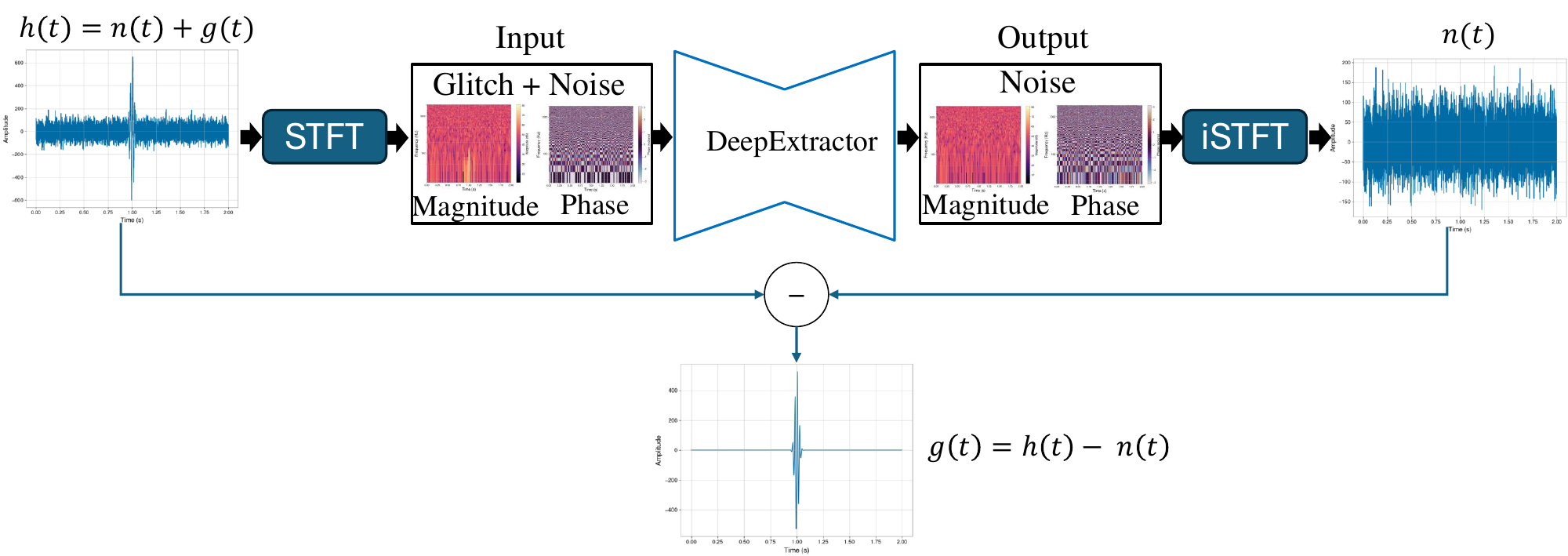}
\caption{An overview of DeepExtractor's reconstruction approach. Two seconds of detector strain $h(t)$ is processed into magnitude and phase spectrogramns using an STFT. This is fed through our network, which maps the instance to the magnitude and phase spectrograms of the underlying background noise. We then use an inverse STFT to yield the corresponding time series $n(t)$. An estimation of the underlying signal or glitch can be simply calculated as $h(t) - n(t)$.}
\label{fig:framework_overview}
\end{figure*}

\subsection{\label{sec:deep_learning_GWs}Deep denoisers for GW data analysis}
Recent studies have explored deep learning for denoising GW signals, focusing on isolating astrophysical events from noise in LIGO and Virgo data with promising results across various scenarios.

Recurrent autoencoders based on Long-Short-Term Memory (LSTM) networks \cite{LSTM, denoising_GWs_Recurrent_AE} have demonstrated robust performance in recovering GW signals. 
Trained on simulated Gaussian noise, these networks excel in real, non-Gaussian LIGO noise, outperforming traditional methods like principal component analysis and dictionary learning, especially at low signal-to-noise ratio (SNR). Notably, the model generalizes well to eccentric GWs outside the training set, showcasing adaptability in diverse signal environments.
Autoencoders using CNNs in the encoder and LSTMs in the decoder have been applied in the AWaRe framework to reconstruct GW events that align with benchmark algorithms. The framework also incorporates uncertainty estimation and can reconstruct signals that overlap with glitches \cite{Chatterjee_DL_BBH_uncertainty, Chatterjee_DL_BBH_overlapping}.

WaveNet architectures \cite{ML_glitch_1} have also been employed for GW denoising in simulated Gaussian and real non-Gaussian, non-stationary LIGO noise, achieving high overlaps ($O > 0.97$) between predicted and numerical relativity waveforms. These networks can additionally extract sine-Gaussian glitches due to structural similarities with GWs while remaining resilient to dissimilar glitches like Gaussian pulses.

U-Net models \cite{Unet_original} have been applied to denoise BBH signals using time-frequency representations \cite{ML_GW_denoising}. Adapted from seismic signal separation studies \cite{seismic_signal_denoise}, this approach leverages the real and imaginary parts of spectrograms to predict separate masks for signal and noise. These masks, applied to input spectrograms, allow reconstruction of the time-domain signal via inverse STFT, yielding strong performance on real astrophysical events.

NNETFIX \cite{NNETFIX_denoise_GWs} addresses signal reconstruction when portions are gated due to glitches. Simulating glitches of varying durations and offsets, this method effectively reconstructs BBH signals with single-interferometer SNRs above $20$, even for glitches lasting hundreds of milliseconds near merger times. This highlights its relevance in the upcoming LIGO-Virgo-KAGRA observing run, where glitch overlap with longer-duration events like BNS signals is expected to increase.

\section{\label{sec:Methods}Methods}

\subsection{\label{sec:UNet_model}DeepExtractor}

\subsubsection{Overview of framework}
\textit{DeepExtractor} is a deep learning framework designed to reconstruct signals or glitches in GW data with power exceeding the underlying detector noise.
An overview of the framework is illustrated in FIG. \ref{fig:framework_overview}.
Built using \textit{PyTorch's} deep learning library \cite{paszke2019pytorch}, \textit{DeepExtractor's} codebase is publicly accessible on GitLab\footnote{\url{https://git.ligo.org/tom.dooney/deepextractor.git}}.
\textit{DeepExtractor} is agnostic to the origin and morphology of excess power in the detector, whether it is a signal or a glitch. 
However, in this paper, we primarily focus on the application of glitch reconstruction.

The framework operates under the first two models of \textit{BayesWave} described in Section \ref{sec:GW_glitches}, where the data contains either an astrophysical signal (signal-only mode) \textbf{or} a glitch (glitch-only mode).
Currently, \textit{DeepExtractor} does not handle the third model assumption, where a glitch overlaps with an astrophysical signal (signal+glitch mode). 
While this capability is critical for a fully-functional glitch mitigation algorithm, this study focuses on the simpler signal-only and glitch-only cases as a foundational step. 
Potential approaches for addressing the signal+glitch scenario are discussed in Section \ref{sec:Conclusions}.

We follow the assumption that a glitch $g(t)$ interacts additively with the strain data in a real detector setting. 
Our goal is to map the input mixture $h(t)$ to the noise component $n(t)$, where $h(t)$ is represented as:
\begin{equation}\label{eq:noise_plus_glitch}
    h(t) = n(t) + g(t)
\end{equation} 
We can then subtract the model output $\hat{n}(t)$ from $h(t)$ to yield an estimate of the underlying signal or glitch embedded in the detector background via $\hat{g}(t) = h(t) - \hat{n}(t)$.

\subsubsection{Training strategy\label{sec:training_strategy}}
The \textit{DeepExtractor} framework utilizes a neural network that directly outputs the background noise component from the data in the spectrogram domain. 
Our core idea is that the distribution of whitened background, which is approximately Gaussian stationary, is easier to model by our network compared to modelling the diverse classes of glitches.

To train \textit{DeepExtractor}, access to the background noise target  \( n(t) \) is required. Therefore, we generate our own synthetic training data. We start with a background noise target \( n(t) \), which represents \( 2\,\text{s} \) of detector data. Next, we inject synthetic glitches \( g(t) \) into \( n(t) \) to create the input mixture \( h(t) \), following Eq. \ref{eq:noise_plus_glitch}. A more detailed discussion on the process of synthesizing the data can be found in Section \ref{sec:simulated_training_data}.

Neural networks can learn in either the time or time-frequency domain, each with trade-offs \cite{compare_time_freq_SE}. \textit{DeepExtractor} is trained in the time-frequency domain using invertible STFTs, which provide rich signal representations that enhance learning \cite{time_vs_frequency_speech, spec_TF_speech_feature_2, spec_TF_speech_feature_4, spec_TF_speech_feature_5}.
While many STFT-based models use only the magnitude spectrogram—assuming identical mixture (input) and source (target) phases \cite{no_phase}—recent work shows that incorporating phase via complex spectrograms improves performance \cite{complex_spectrograms_denoise, complex_spec_CNN, phase_processing}.
Accordingly, we model both magnitude and phase components of the complex STFT spectrograms to enable accurate time-domain reconstruction of signals and glitches.

To provide the inputs and targets, we first transform the time-domain signals \( h(t) \) and \( n(t) \) into the time-frequency domain by computing their complex spectrograms, \( h(t, f) \) and \( n(t, f) \), using the STFT (see Section \ref{sec:simulated_training_data}).
The spectrograms are then decomposed into magnitude and phase components, each with dimension $\mathbb{R}^{h \times w}$. These are then processed in two separate input and output channels of the network, using a data structure of dimension $\mathbb{R}^{2\times h \times w}$ (see FIG. \ref{fig:UNet_architecture}).
Thus, \textit{DeepExtractor} takes the magnitude and phase components of \( h(t, f) \) as input and outputs the corresponding components of the estimated noise \( \hat{n}(t, f) \).

We optimize the network by minimizing a mean squared error (MSE) loss function, which compares the predicted components \( \hat{n}(t, f) \) with the true background noise components \( n(t, f) \), as defined below:
\begin{equation}
    \frac{1}{N}\sum_{i=1}^N \left( \hat{n}_i(t, f) - n_i(t, f) \right)^2,
\end{equation}
where $i$ is the index of the data sample and $N$ is the batch size.

During inference, the predicted magnitude and phase components of \( \hat{n}(t, f) \) are recombined into a complex spectrogram, which is then transformed back to the time domain using the inverse STFT to produce the noise estimate \( \hat{n}(t) \). By subtracting \( \hat{n}(t) \) from the input mixture \( h(t) \), the framework isolates the underlying signal or glitch, \( \hat{g}(t) \). This approach ensures the network focuses on learning the simpler distribution of Gaussian noise, avoiding the complexities associated with directly modeling the diverse combinations of glitches present in the data.

\textit{DeepExtractor} employs the \textit{Adam} optimizer \cite{ADAM} with a batch size of 32. Losses for \textit{DeepExtractor} and other time-frequency models are calculated directly on spectrogram data, while for time-domain benchmarks, they are computed on raw time-series data. To optimize the learning process, the training incorporates the \texttt{ReduceLROnPlateau} scheduler provided by \textit{Pytorch}, which reduces the learning rate after four epochs without improvement, and early stopping is applied after ten such epochs. This ensures an efficient yet flexible training duration tailored to the model's performance on validation data.

All models in this study were trained on an NVIDIA A100 GPU (80GB memory). 
Our best-performing \textit{DeepExtractor} configuration trained on the simulated dataset (see Section \ref{sec:simulated_training_data}) in approximately 8 hours, converging after 24 epochs. 
The subsequent transfer learning phase on real LIGO data (see Section \ref{sec:real_backgrounds}) required only 3 additional epochs to converge and completed in under 20 minutes.

\subsubsection{U-Net Architecture}
\begin{figure}[t]
\centering
\includegraphics[width =\linewidth]{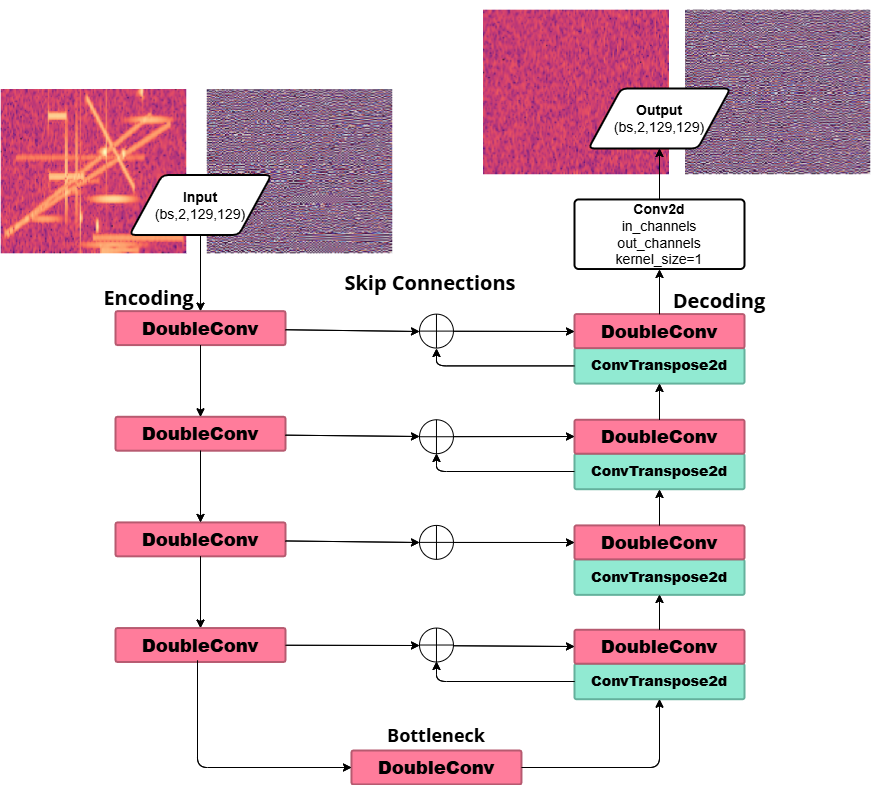}
\caption{The U-Net architecture featured in \textit{DeepExtractor} applied to batches (bs) of STFT data ($\mathbb{R}^{2\times h \times w}$), illustrating its characteristic ‘U’-shaped structure. The network processes both the magnitude and phase components of the STFT simultaneously through two input and output channels.}
\label{fig:UNet_architecture}
\end{figure}

\textit{DeepExtractor} employs a U-Net architecture \cite{Unet_original} for its ability to capture both high- and low-level features through skip connections. These connections concatenate encoder features with their corresponding decoder features (FIG. \ref{fig:UNet_architecture}), preserving fine-grained details while supporting a large receptive field. This design has made U-Nets a state-of-the-art choice for signal denoising \cite{Wave_UNET}, acoustic source separation \cite{U-Net_denoise, speech_time_domain_1}, and image denoising \cite{UNet_image}.
U-Nets are particularly effective for spectrogram-based representations of signals and glitches spanning broad time-frequency ranges. Unlike deep autoencoders, which may lose information through aggressive compression, U-Nets preserve structural details throughout the network, enabling more accurate and faithful reconstructions \cite{learnable_WPT_denoising}. \\

\textit{DeepExtractor} is a 2D model designed to process magnitude and phase spectrograms. 
It accepts two input channels and outputs two spectrograms of the same dimensionality, as illustrated in FIG. \ref{fig:UNet_architecture}.\\

\noindent \textbf{Encoder (Downsampling Path):} The encoder extracts hierarchical features from the input through:
\begin{itemize}
    \item \textbf{DoubleConvolution Blocks:} Each block consists of two 2D convolutional layers (kernel size 3), each followed by batch normalization \cite{batch_norm} and ReLU activation \cite{RELU}, to enhance feature extraction while preserving spatial resolution.
    \item \textbf{MaxPooling:} A 2D max-pooling operation (kernel size 2) reduces spatial resolution by half after each DoubleConvolution block, enabling hierarchical feature representation.
\end{itemize}
Feature maps from the encoder are stored as skip connections for the decoder.\\

\noindent \textbf{Decoder (Upsampling Path):} The decoder reconstructs the original input resolution while integrating features from the encoder:
\begin{itemize}
    \item \textbf{Transposed Convolutions:} Upsampling is performed using transposed convolutional layers that double the spatial resolution.
    \item \textbf{Skip Connections:} Features from the encoder at corresponding resolutions are concatenated with the upsampled features to preserve fine details.
    \item \textbf{DoubleConvolution Blocks:} Combined features are processed through DoubleConvolution blocks for refinement.
\end{itemize}

\noindent \textbf{Bottleneck:} At the deepest level, the bottleneck processes the feature map using a DoubleConvolution block, capturing the most abstract representations of the input. \\

\noindent \textbf{Output:} The model concludes with a single 2D convolutional layer (kernel size 1) with a linear activation function, producing outputs with the desired dimensionality. \\

\noindent \textbf{Time-Domain Variant:} We also experiment using a 1D time-domain counterpart of \textit{DeepExtractor's} U-Net architecture (UNET1D). 
This replaces all 2D operations with 1D equivalents and processes a single input/output channel. Specifically, the 1D model maps the raw, whitened time series $h(t)$ to the noise component $n(t)$, omitting the STFT and iSTFT transformations used in the 2D variant, as shown in FIG. \ref{fig:framework_overview}.

\subsection{\label{sec:Training_data}Datasets}
\begin{figure*}[t]
\centering
\begin{subfigure}{0.2\textwidth}
  \centering
  \includegraphics[width = \textwidth]{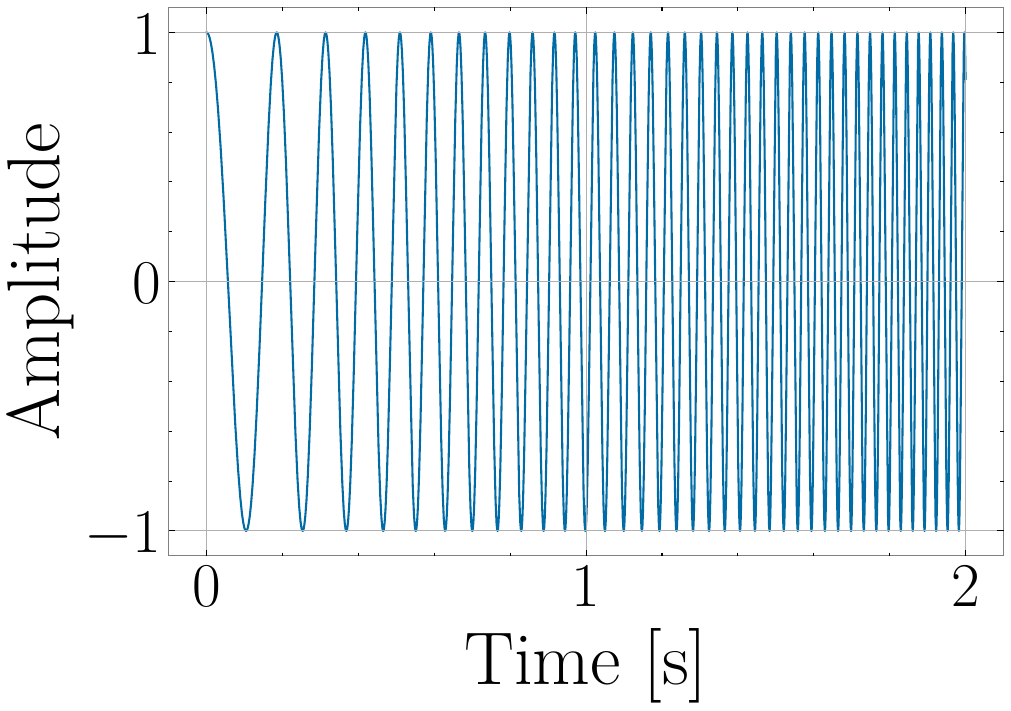}
  \caption{Chirp}
  \label{fig:chirp_signal}
\end{subfigure}%
\begin{subfigure}{0.2\textwidth}
  \centering
  \includegraphics[width = \textwidth]{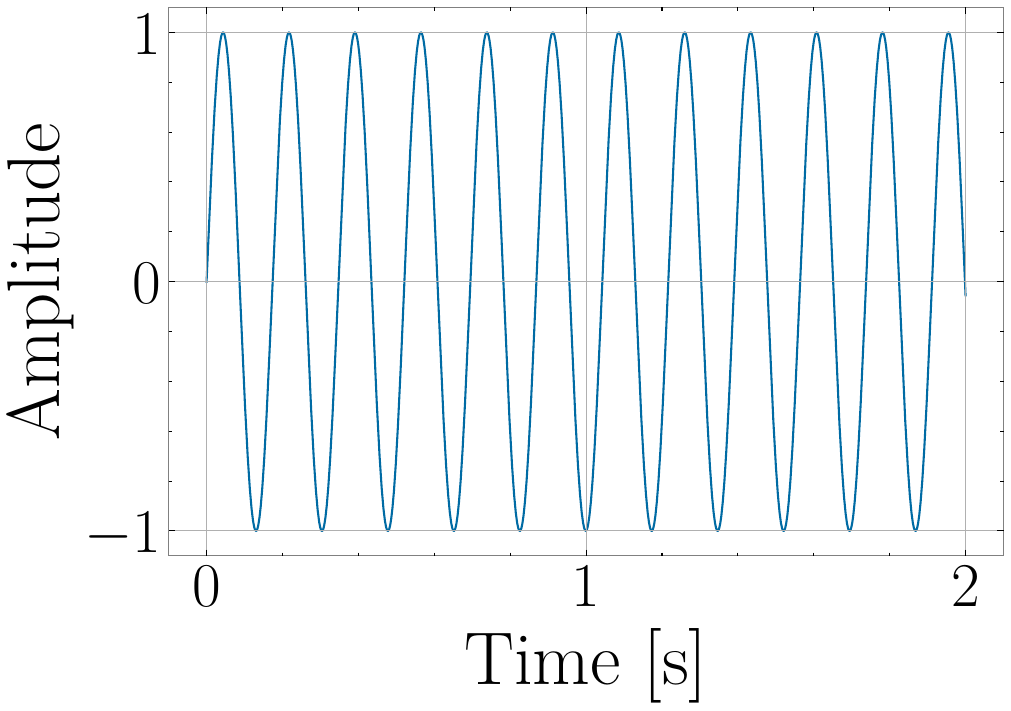}
  \caption{Sine}
  \label{fig:PCA_real_vs_fake_all}
\end{subfigure}%
\begin{subfigure}{0.2\textwidth}
  \centering
  \includegraphics[width = \textwidth]{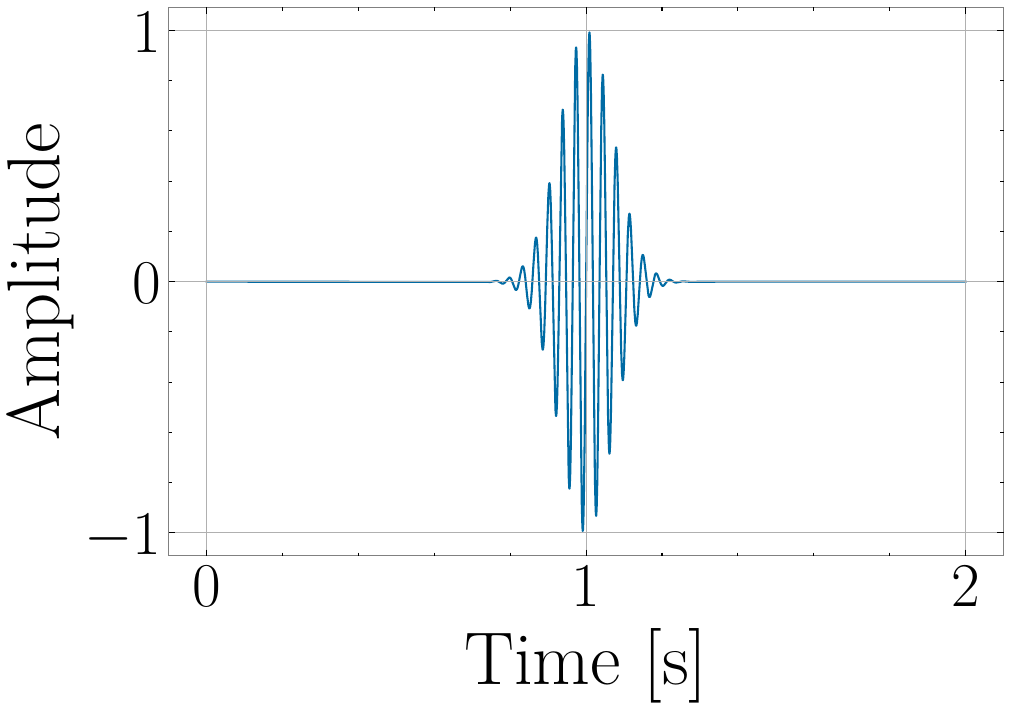}
  \caption{Sine-Gaussian}
  \label{fig:sine_gaussian_signal}
\end{subfigure}%
\begin{subfigure}{0.2\textwidth}
  \centering
  \includegraphics[width = \textwidth]{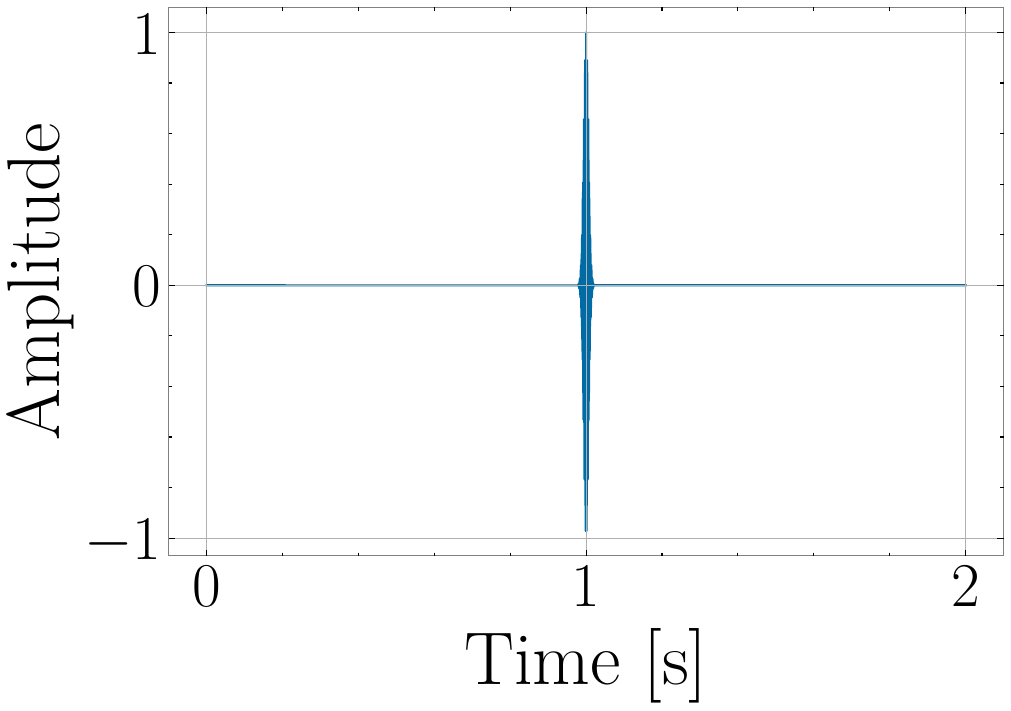}
  \caption{Gausian Pulse}
  \label{fig:gaussian_pulse_signal}
\end{subfigure}%
\begin{subfigure}{0.2\textwidth}
  \centering
  \includegraphics[width = \textwidth]{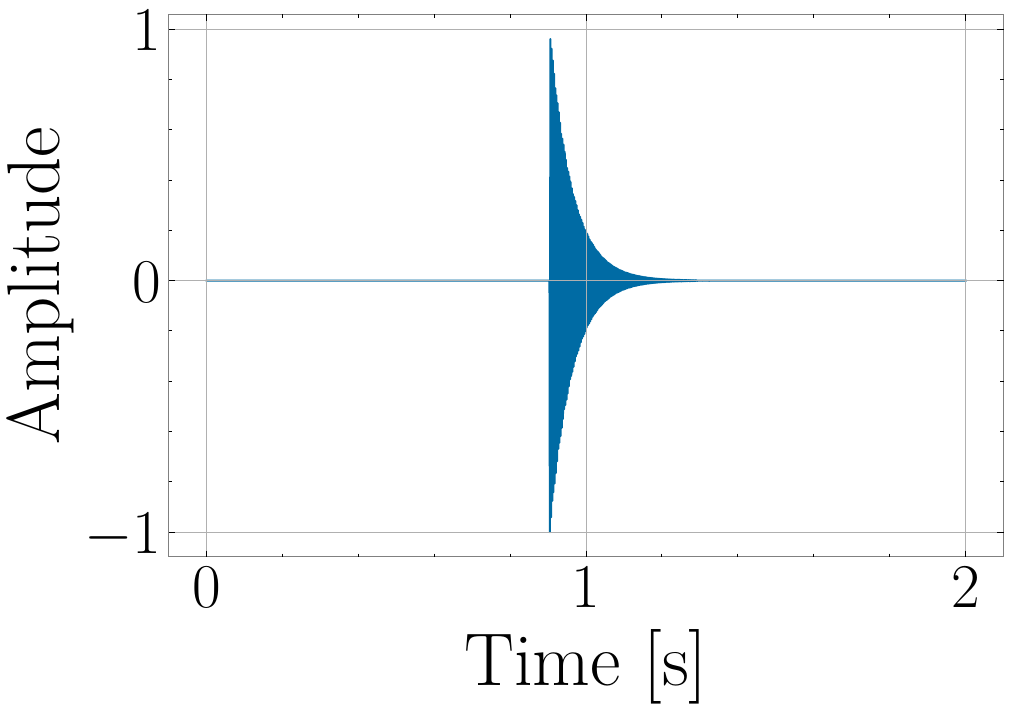}
  \caption{Ringdown}
  \label{fig:ringdown_signal}
\end{subfigure}%
\caption{Examples of each of the five training glitch classes; chirp, sine, sine-gaussian, gaussian pulse and ringdown. For each $2\,$s training sample, anywhere between 1 and 30 of these signals (selected randomly) are injected into the Gaussian background sample.}
\label{fig:training_examples}
\end{figure*}
\subsubsection{\label{sec:simulated_training_data}Simulated data}

To properly validate our model, we first implement a fully simulated approach for both training and validation, simulating the background noise and glitches. 
We simulate white noise training targets \( n(t) \) with a duration of 2 seconds at a sampling rate of $f_s = 4,096\,$ Hz, yielding 8,092 data points. 
We generate 250,000 background samples for training and 25,000 for validation.

To create the corresponding input samples, we inject various analytical waveforms, represented by 
$g(t)$, into the background noise $n(t)$, as per Eq. \ref{eq:noise_plus_glitch}.
While \textit{BayesWave} models glitches exclusively as linear combinations of sine-Gaussian wavelets, our approach broadens the space of basis waveforms to improve the network’s generalization capability. 
Specifically, we draw from five proxy glitch classes—\textit{chirp}, \textit{sine}, \textit{sine-Gaussian}, \textit{Gaussian pulse}, and \textit{ringdown}—illustrated in FIG. \ref{fig:training_examples}.

The \textit{chirp} class was selected for its resemblance to the frequency evolution characteristic of CBC signals. 
The \textit{sine} class captures narrow-band features across varying frequencies. 
The \textit{sine-Gaussian} class, inspired by \textit{BayesWave}, along with the \textit{Gaussian pulse} and \textit{ringdown} classes, serves as an analytical proxy for gravitational-wave signals expected from a variety of burst sources \cite{All_sky_burst}. 
Since these waveforms are commonly used in unmodelled burst searches, we investigate their suitability for unmodelled glitch reconstruction in our \textit{DeepExtractor} framework.

Each waveform class has its own parameters that are randomly sampled during generation, such as frequency range (from $1\,$ Hz to the Nyquist frequency of $2,048\,$Hz), bandwidth, and duration (randomly chosen between $0.125\,$s and $2\,$s).
The lower bound of $0.125\,$s was chosen since over 90\% of high-confidence \textit{Gravity Spy} glitches exceed this duration. 
Additionally, the analytical waveforms used for training often contain features—such as individual sine-Gaussian cycles—that evolve on timescales shorter than $0.125\,$s.
While shorter durations could be considered, this threshold is expected to capture the majority of short-duration structures relevant for glitch reconstruction. 
For more information on these training classes, please see Appendix \ref{sec:simulated_glitch_descriptions} and our code repository.
The SNR is varied between 1 and 250, scaled according to

\begin{equation}\label{eq:snr_scaling}
    \rho_{opt}^2 = 4\int^{f_{high}}_{f_{low}}\frac{|g(f)|^2}{S_n(f)}df
\end{equation}

where $g(f)$ and $S_n(f)$ represent the Fourier transform of the injected waveform and the detector noise PSD, respectively \cite{Allen_2012_findchirp}. We set $S_n(f)$ to unity for convenience, since we are working with simulated whitened detector noise with a flat PSD. 

To improve generalization and avoid overfitting, we inject linear combinations of glitches into each background sample, where:

\begin{equation}\label{eq:glitch_sum} \mathbf{g} = \sum_{i=1}^kg_i(t) \end{equation}

and $k \in [1, 30]$. Futhermore, each $g_i(t)$, with $i \in [1, k]$, is randomly time-shifted within the $2\,$s duration. This procedure produces a diverse set of composite proxy glitches, which may or may not overlap in time and frequency.
In the context of the training data, $g(t)$ in Eq. \ref{eq:noise_plus_glitch} can be replaced with $\mathbf{g}$ to reflect this linear combination.
Inspired by \textit{BayesWave}'s approach of modeling glitches as sums of sine-Gaussian wavelets, our setup injects between 1 and 30 glitches per sample, drawn from a broader five-class proxy glitch space.

Additionally, glitches can be scaled to low SNRs, so some training samples include injections with very little (or even negligible) contribution to the data.
We hypothesize that this approach helps the model to learn features inherent to the background noise when glitches are not present in the data.
We test this hypothesis on both simulated (Section \ref{sec:Simulated_experiments}) and real detector noise (Section \ref{sec:GravitySpy_experiments}). 

To prepare the data for training \textit{DeepExtractor}, we scale (standardize) the time-domain inputs and outputs using \texttt{StandardScaler} from \textit{sklearn} \cite{scikit_learn}, which transforms the data so that it has a mean of 0 and a standard deviation of 1.
This standardization method assumes a Gaussian distribution of the data and is appropriate given the Gaussian-like characteristics of the background noise.

We fit the scaler exclusively to the model input (i.e., the glitch-injected samples) and use this input-fitted scaler to transform both the input and the background noise target. This ensures that both inputs and outputs occupy the same standardized space.
While this data must be then transformed using the STFT to train \textit{DeepExtractor}, it is this preprocessed data that is used to train the time-domain benchmarks (eg. UNET1D).\\

\begin{figure}[t]
\centering
\begin{subfigure}[t]{0.48\linewidth}
  \centering
  \includegraphics[width = \linewidth]{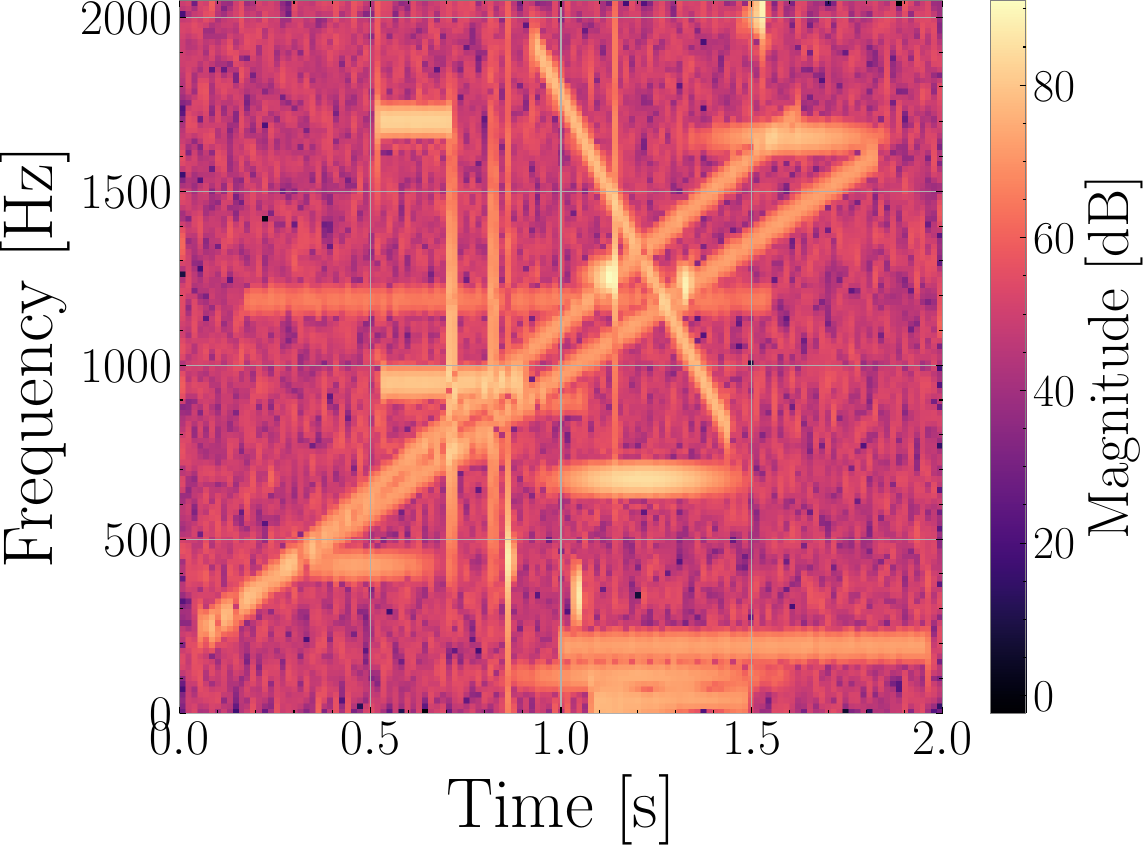}
  \label{fig:glitch_stft_mag}
\end{subfigure}%
\begin{subfigure}[t]{0.48\linewidth}
  \centering
  \includegraphics[width = \linewidth]{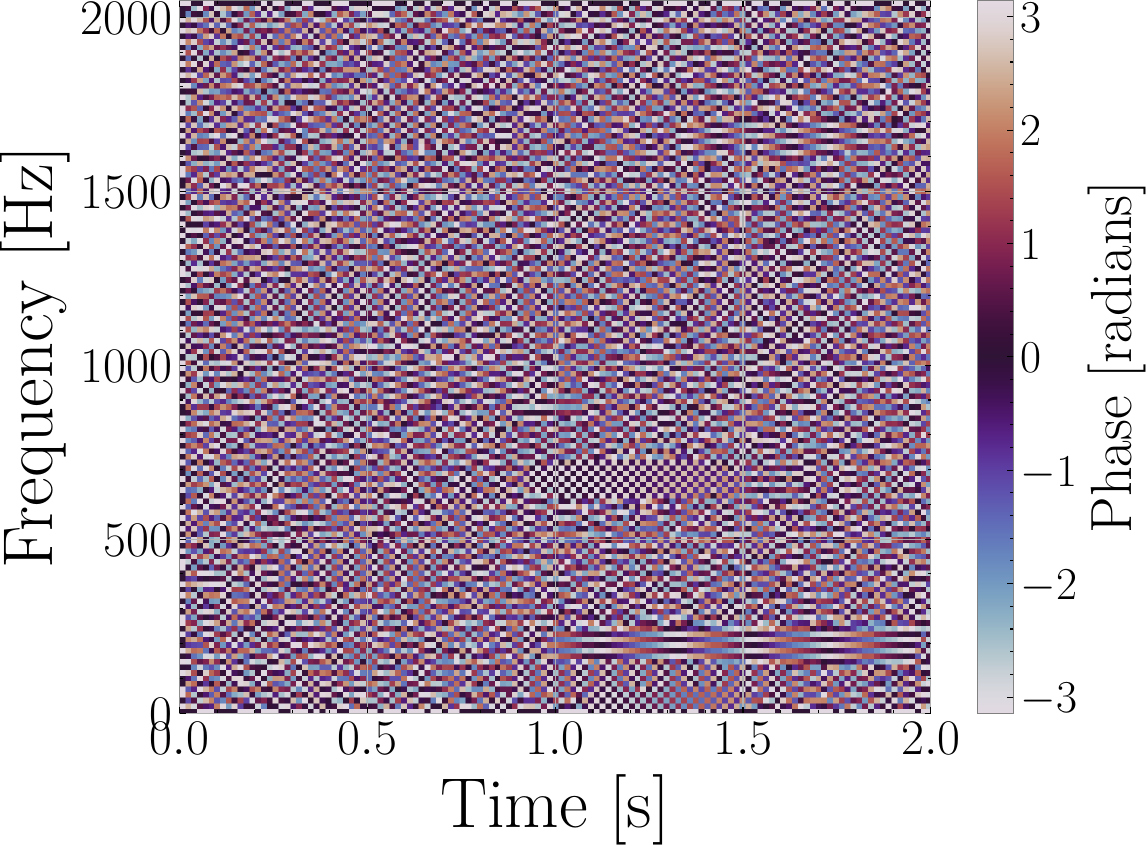}
  \label{fig:glitch_stft_phase}
\end{subfigure}%

\begin{subfigure}[t]{0.48\linewidth}
  \centering
  \includegraphics[width = \linewidth]{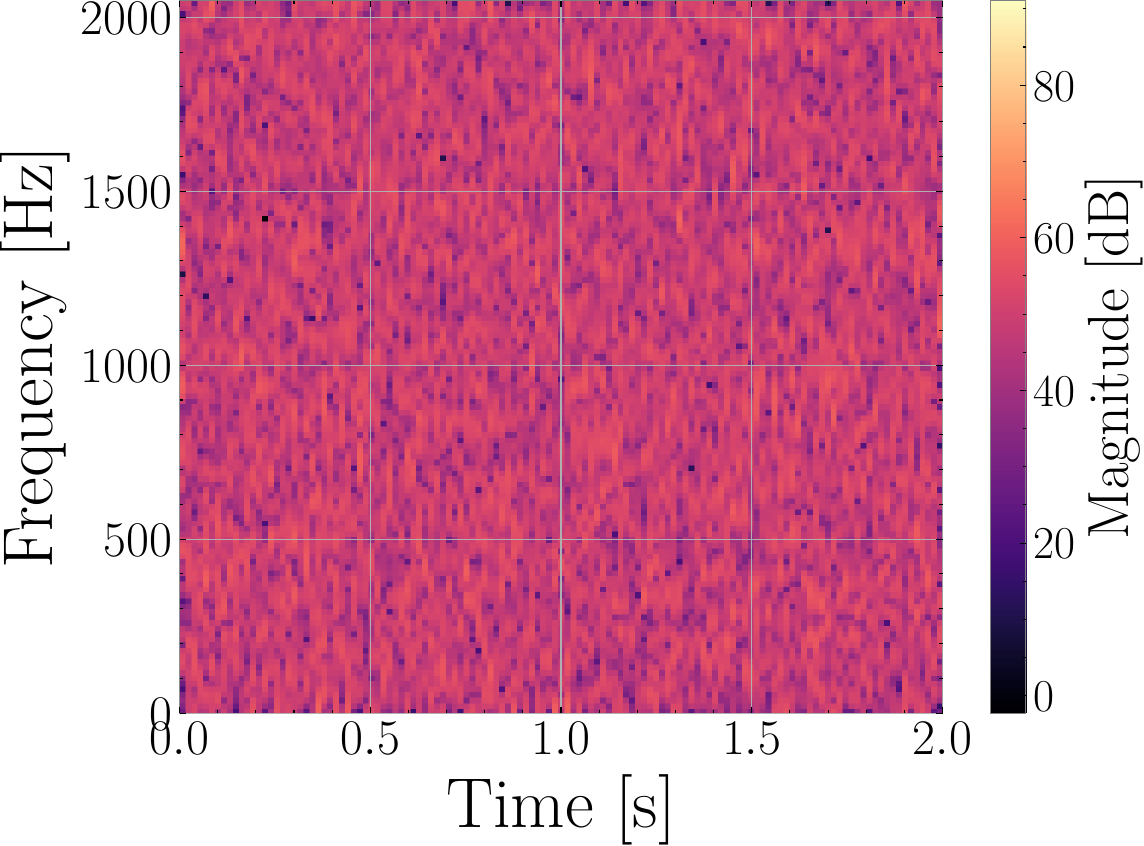}
  \caption{Magnitude}
  \label{fig:background_stft_mag}
\end{subfigure}%
\begin{subfigure}[t]{0.48\linewidth}
  \centering
  \includegraphics[width = \linewidth]{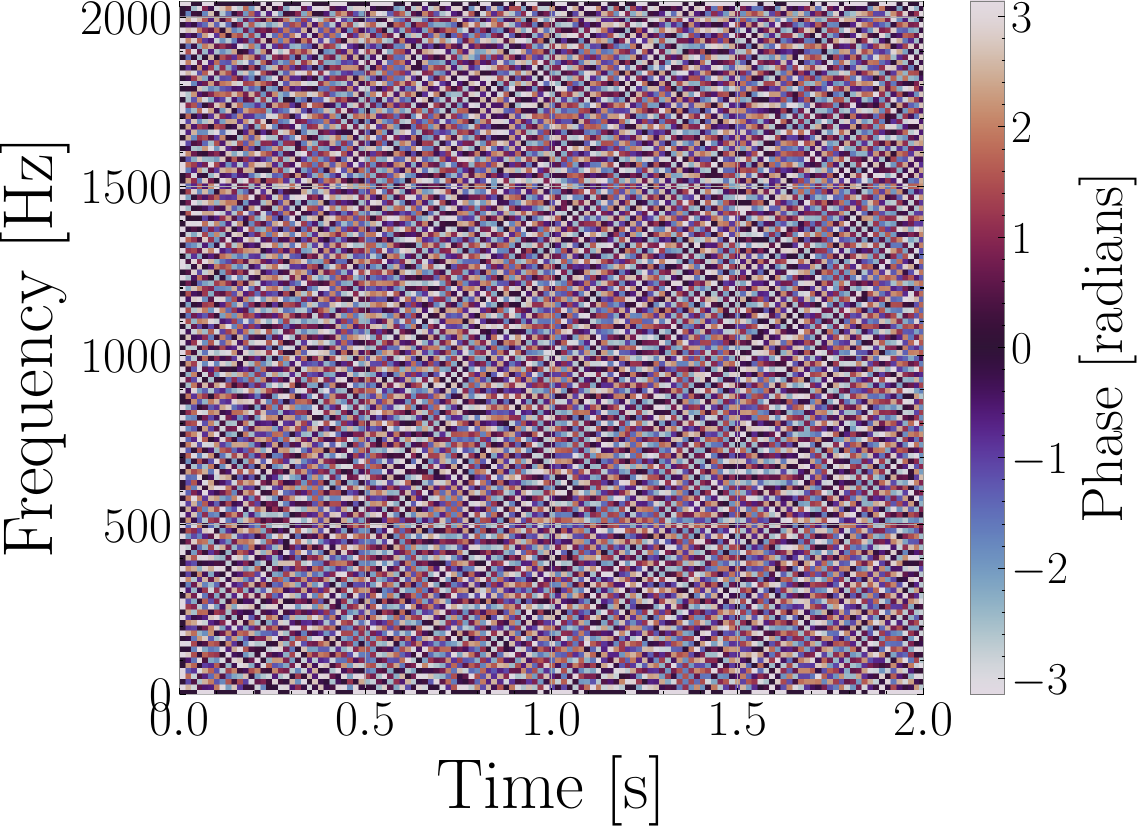}
  \caption{Phase}
  \label{fig:background_stft_phase}
\end{subfigure}%
\caption{Magnitude and phase STFT spectrograms for noise+glitch input (top) and noise-only target (bottom) used to train \textit{DeepExtractor}.}
\label{fig:STFT_mag_phase}
\end{figure}

\noindent \textbf{Short time Fourier transform (STFT):} As mentioned in Section \ref{sec:training_strategy}, \textit{DeepExtractor} and the 2D benchmarks are trained on spectrograms calculated from the above preprocessed data using the STFT. 
Since the time-domain data is already standardized, no additional scaling is performed on the spectrograms.
FIG. \ref{fig:STFT_mag_phase} shows examples of the magnitude and phase components for both the input and target spectrograms.

We choose STFT parameters to ensure compliance with the Constant Overlap-Add (COLA) condition \cite{COLA} (please see Appendix \ref{sec:STFT_parameters} for more information). 
A Hann window is employed to minimize spectral leakage, further improving the quality of the spectrogram representation \cite{spectral_leakage}.
These chosen settings yield complex spectrograms with dimensions of $257\times257$, which we separate into magnitude and phase spectrograms. 
To investigate the effect of varying time and frequency resolutions on signal and glitch reconstruction, we also experiment with alternate parameter configurations that produce spectrograms of dimensions $129\times129$ and $65\times129$ while ensuring that all configurations satisfy the COLA condition.

\subsubsection{\label{sec:real_backgrounds}Real LIGO O3 backgrounds}

\begin{figure}[t]
\centering
\begin{subfigure}[t]{\linewidth}
  \centering
  \includegraphics[width = \linewidth]{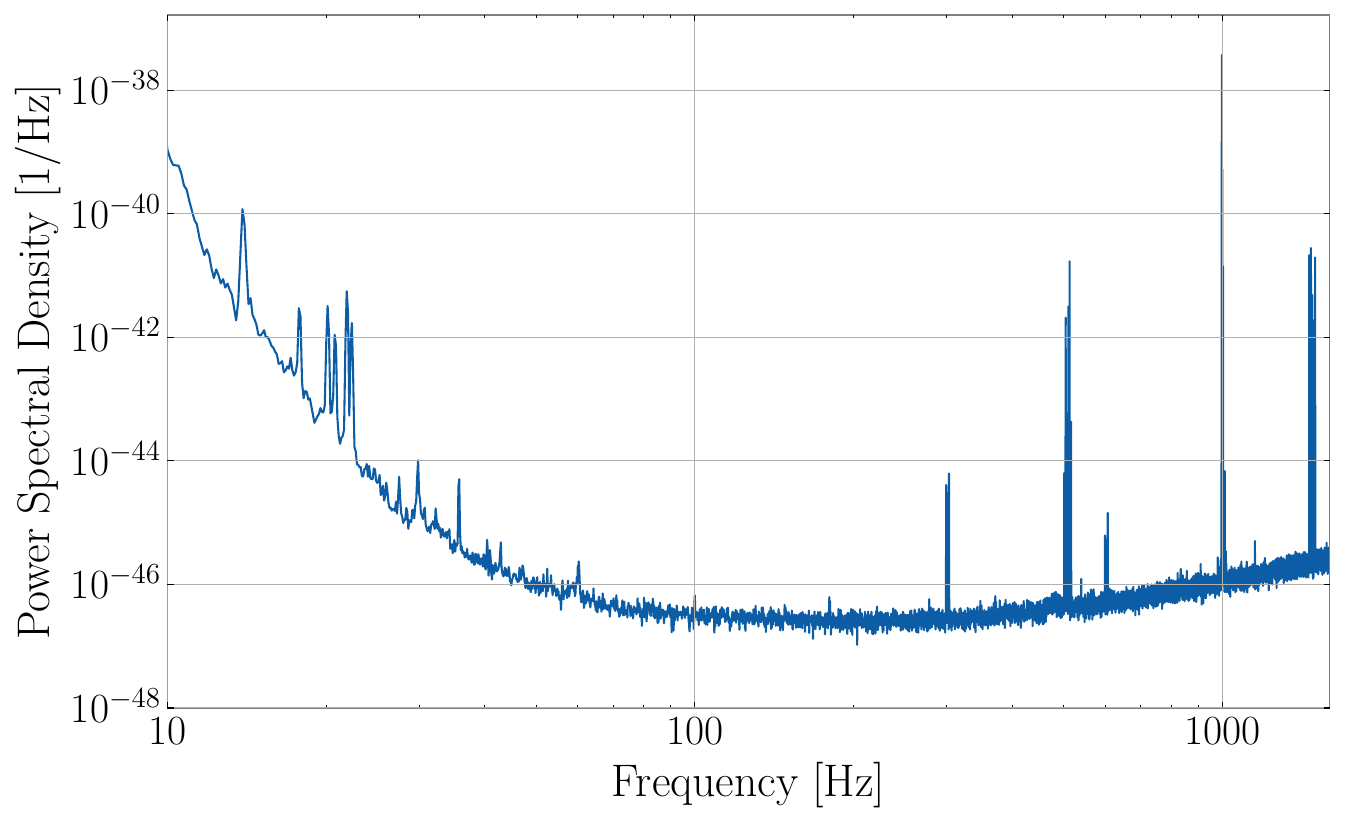}
  \caption{}
  \label{fig:psd_LIGO}
\end{subfigure}%

\begin{subfigure}[t]{\linewidth}
  \centering
  \includegraphics[width = \linewidth]{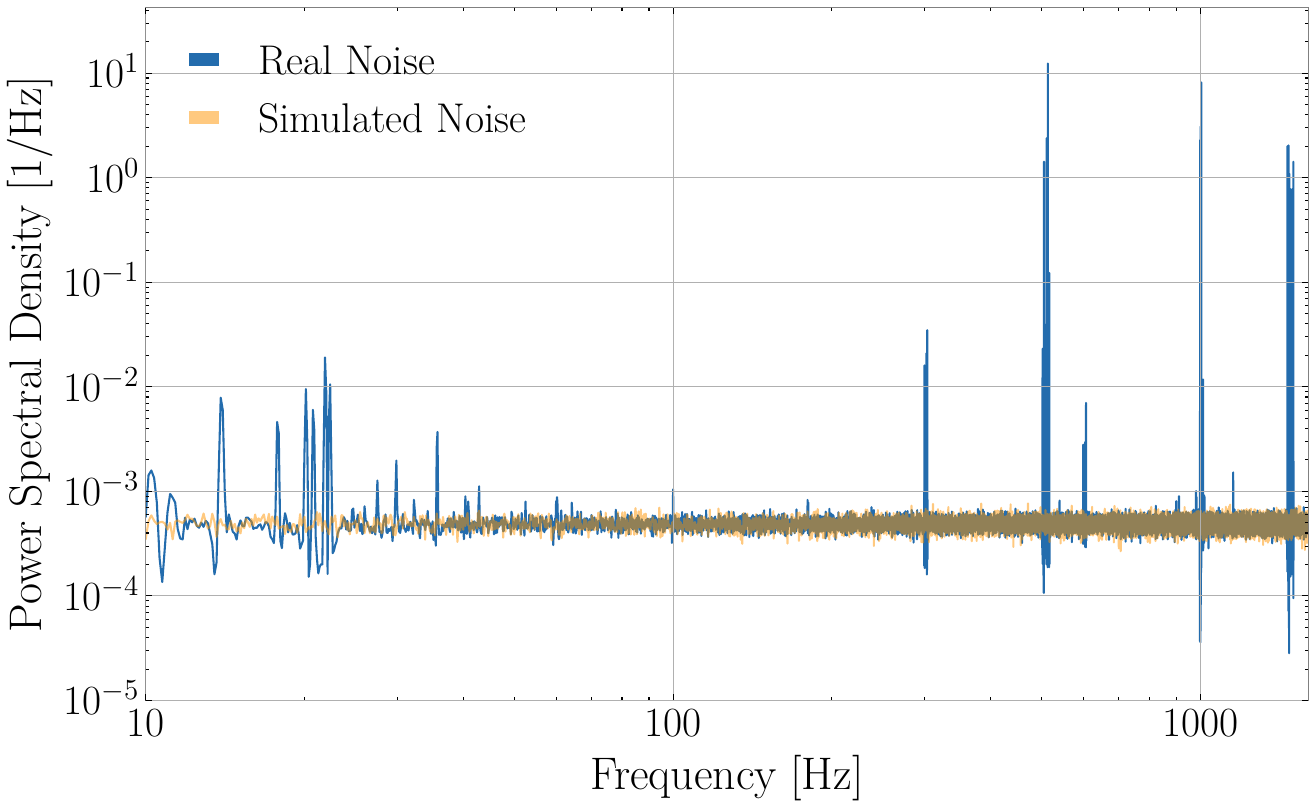}
  \caption{}
  \label{fig:whitened_psd}
\end{subfigure}%
\caption{Power spectral density (PSD) of \textbf{(a)} real LIGO Hanford detector noise and \textbf{(b)} the same PSD whitened and our simulated white noise. In comparison, it is visible that our simulated noise misses instrumental or environmental lines commonly found in the GW detectors.}
\label{fig:psd_comparison}
\end{figure}

GW detector noise is often modeled as Gaussian and stationary over short timescales, but it is generally non-Gaussian and exhibits a time-varying PSD. The sensitivity of GW detectors is constrained by broadband noise sources, such as seismic noise at low frequencies and quantum sensing noise at high frequencies \cite{detchar_transient}. An example PSD from LIGO Hanford is shown in FIG. \ref{fig:psd_comparison}. The bottom plot compares the PSD of whitened data using Welch averaging \cite{welch} to that of simulated white noise described in the previous section. While whitening reduces broadband noise, narrow-band spectral lines—caused by electrical and mechanical sources or resonances—persist \cite{spectral_lines}. In contrast, the simulated noise PSD shows a smooth, flat trend with no spectral lines, underscoring the complexities of real GW detector noise.

To accurately reconstruct signals and glitches in real GW data, it is essential to account for these unique noise features. A model trained exclusively on simulated data will inadvertently suppress spectral lines when applied to real data, leading to contaminated reconstructions of both signals and glitches. Therefore, incorporating real data into the training of \textit{DeepExtractor} is crucial for adapting to the non-Gaussian, non-stationary nature of GW detector noise and ensuring accurate modeling of real events outside the training distribution.

To address this challenge, we apply transfer learning \cite{transfer_learning} to our \textit{DeepExtractor} model, leveraging knowledge learned from simulated data to enhance performance or reduce training effort on real data. Transfer learning allows a model trained for one task to adapt to a related task, making it an effective strategy in this context.

We restrict the scope of this study to LIGO data during O3. 
We accumulate real backgrounds by identifying GPS times for all GW events and \textit{Omicron} triggers. Background segments of $14\,$s are extracted from \textit{GWOSC} \cite{GWOSC_cite} data during O3a and O3b, ensuring these segments are at least $2\,$s away from any trigger. 
These segments are whitened using Welch’s method \cite{welch}, with the first and last $2\,$s discarded to mitigate boundary artifacts. 
The resulting $10\,$s segments are divided into $2\,$s samples.
We perform a final validation step using Q-scans for each $2\,\mathrm{s}$ sample, discarding any sample with a maximum pixel (tile) energy — defined as the squared magnitude of the Q-transform coefficient for a time-frequency tile — exceeding 20 (see \cite{Q_scan_pixel_energy} for details on pixel energy). 
This ensures that no sample contains excessive power that could interfere with the training of \textit{DeepExtractor}. 
After applying this criterion, we collect a total of $45,400$ background samples from LIGO's O3 run, distributed as follows: $18,400$ from Hanford O3a, $1,300$ from Livingston O3a, $18,600$ from Hanford O3b, and $7,100$ from Livingston O3b\footnote{While Livingston has nearly twice as many Omicron triggers as Hanford, a significant portion of Livingston background samples were also excluded during the Q-scan validation due to persistent excess power, even outside the Omicron triggers.}.
For injections, we follow the approach described in Section \ref{sec:simulated_training_data}, with the key distinction that the injections are now added to real, whitened O3 backgrounds.

\subsubsection{\label{sec:test_datasets}Unseen Test Classes}

To evaluate the generalization performance of \textit{DeepExtractor}, we utilize two glitch generators: \textit{gengli} and cDVGAN \cite{GENGLI, cdvgan}. These generative deep learning models produce realistic glitches, also enabling robust testing on unseen glitch classes.
Trained on \textit{blip}\footnote{https://gswiki.ischool.syr.edu/en/Glitches/Blip} glitches extracted from LIGO O2 data using \textit{BayesWave}, \textit{gengli} generates reliable representations of this specific glitch type in both Hanford and Livingston detectors separately.
A conditional GAN trained on preprocessed LIGO O3 data, cDVGAN produces \textit{blip} and \textit{tomte}\footnote{https://gswiki.ischool.syr.edu/en/Glitches/Tomte} glitches along with BBH signals with component masses in the range of $[30, 160]M_{\odot}$, a spin of zero and fixing $m_1>m_2$. 
As a conditional model, cDVGAN can create hybrid samples (\textit{simplex} and \textit{uniform}) that blend features across all three training classes, introducing even greater diversity. 
By leveraging the class-mixing feature of cDVGAN, we create a diverse test dataset to assess \textit{DeepExtractor}'s ability to generalize to unseen classes.
Examples of \textit{gengli} and \textit{cDVGAN} generations can be found in Appendix \ref{sec:cDVGAN generations}.

\subsection{\label{sec:Experiments}Experiments}

\subsubsection{\label{sec:Simulated_experiments}Simulated Experiments}
We evaluate \textit{DeepExtractor}'s performance by injecting simulated glitch samples into simulated background noise. This controlled environment allows for a direct comparison between the extracted glitches and their ground-truth counterparts, helping identify the most effective data representations in terms of dimensionality and domain (time vs. time-frequency).

Simulated glitches are generated using both analytical waveforms and samples from the glitch generators discussed in the previous section.
While the analytical waveforms represent known classes from training, we include unseen classes using glitches simulated with \textit{gengli} and \textit{cDVGAN}. 
Samples are injected into simulated white noise at SNRs $ \in [7.5, 100]$ using Eq. \ref{eq:snr_scaling}.
We limit the maximum SNR of the test glitches to 100 (reduced from 250 in our training set) to create a controlled environment for glitch reconstruction. Furthermore, over 90\% of high-confidence glitches\footnote{Those classified with a Gravity Spy confidence of at least 90\%.} observed during O3 had an SNR below 100.

The results are evaluated using the mismatch metric, defined as

\begin{equation}\label{eq:mismatch}
    \mathcal{M}(g_1, g_2) = 1 - M(g_1, g_2)
\end{equation}

where

\begin{equation}
    M(g_1, g_2) = \max_{\phi_c, t_c} \frac{\langle g_1 | g_2 \rangle}{\sqrt{\langle g_1 | g_1 \rangle \langle g_2 | g_2 \rangle}}
\end{equation}

with the inner product given by:

\begin{equation}
    \langle g_1 | g_2 \rangle = 4 \, \Re \int_{f_{\text{low}}}^{f_{\text{high}}} \frac{\tilde{g}_1(f) \tilde{g}_2^*(f)}{S_n(f)} \, \mathrm{d}f
\end{equation}

and the normalization:

\begin{equation}
    \langle g | g \rangle = 4 \, \int_{f_{\text{low}}}^{f_{\text{high}}} \frac{|\tilde{g}(f)|^2}{S_n(f)} \, \mathrm{d}f
\end{equation}

Here, $\tilde{g}_1(f)$ and $\tilde{g}_2(f)$ are the Fourier transforms of the waveforms ($*$ represents complex conjugate), $S_n(f)$ is the PSD of the noise and $f_{\text{low}}$ and $f_{\text{high}}$ are the frequency bounds of the analysis.
The match is maximized over relative shifts in time $t_c$ and phase $\phi_c$.

We evaluate the performance of \textit{DeepExtractor} by comparing it against several deep learning benchmarks. These include a 1D denoising CNN (DnCNN1D) \cite{CNN_img_denoise} and 1D and 2D autoencoders \cite{denoising_AEs} that are structurally similar to the U-Net, but without skip connections. 
We also assess the performance of the time-domain variant of \textit{DeepExtractor's} U-Net architecture (UNET1D) to compare its effectiveness across domains.
This comparison helps evaluate \textit{DeepExtractor}'s performance relative to other denoising and feature extraction architectures.
Furthermore, we examine the performance of both \textit{DeepExtractor} and UNET1D when directly mapping to $g(t)$, investigating the effectiveness of our approach in first mapping to $n(t)$ and subtracting it to yield $\hat{g}(t)$.
Finally, we apply \textit{DeepExtractor} to background samples without injections, where the ideal outcome is for the output to match the input, and the inferred glitch component $\hat{g}(t)$ to be zero.

We explore additional target mappings for UNET1D by conducting several ablation studies. An ablation study in machine learning is a systematic experimentation technique used to understand the contribution of individual components (target mappings) of a model to its overall performance.\\

\noindent \textbf{Dual-Channel Output:} We evaluate a dual-channel output configuration where the network predicts $\hat{n}(t)$ in one channel and $\hat{\mathbf{g}}(t)$ in the other, using both $n(t)$ and $\mathbf{g}(t)$ as targets. 
An extra loss ensures the sum of outputs equals the mixture input ($\hat{\mathbf{g}}(t) + \hat{n}(t) == h(t)$), enforcing consistency with Eq. \ref{eq:noise_plus_glitch}.\\

\noindent \textbf{Difference Output Layer:} In a variation of the dual-channel setup, we omit the explicit $\mathbf{g}(t)$ target. Instead, the second channel predicts $\hat{\mathbf{g}}(t)$ indirectly as $\hat{\mathbf{g}}(t) = h(t) - \hat{n}(t)$.
This approach has been successfully applied in speech enhancement and audio source separation tasks \cite{wavenet_speech, Wave_UNET}.

\subsubsection{\label{sec:BayesWave_comparison}BayesWave Comparison}
This experiment compares \textit{DeepExtractor} with \textit{BayesWave}, the current state-of-the-art method for glitch mitigation in GW physics. Following the identification of the optimal \textit{DeepExtractor} configuration in Section \ref{sec:Simulated_experiments}, we conduct a controlled evaluation of both methods using simulated glitch injections.

In this experiment, we inject 30 samples from each of the 12 glitch classes mentioned in Table \ref{tab:match_table} (360 samples in total) into background noise at SNRs $ \in [15, 100]$, then process the samples using both \textit{DeepExtractor} and \textit{BayesWave} for a direct, 1:1 comparison.
We increase the lower SNR bound from $7.5$ (as used in Section \ref{sec:Simulated_experiments}) to $15$ to prevent potential convergence issues with \textit{BayesWave}.
The performance of each method is assessed by evaluating the mismatch (Eq. \ref{eq:mismatch}) between the injected and reconstructed samples, following the same approach outlined in the previous section.

\begin{table*}
\centering
\setlength{\tabcolsep}{5pt}
\resizebox{\textwidth}{!}{
\begin{tabular}{lccccccc}
    \toprule
    Glitch Type & UNET1D & DnCNN1D & Autoencoder1D & Autoencoder2D & DeepEx. ($65\text{x}129$) & DeepEx. ($129^2$) & DeepEx. ($257^2$) \\
    \midrule
    chirp & $2.0^{+2.1}_{-0.8}$ & $26.1^{+20.8}_{-13.9}$ & $2.3^{+2.4}_{-0.9}$ & $41.2^{+20.1}_{-9.3}$ & $1.9^{+1.7}_{-0.6}$ & $1.1^{+0.9}_{-0.4}$ & $\mathbf{0.8}^{+0.7}_{-0.3}$ \\
    sine & $1.7^{+2.8}_{-0.6}$ & $23.5^{+27.6}_{-11.5}$ & $1.9^{+3.7}_{-0.7}$ & $41.1^{+20.2}_{-9.7}$ & $1.7^{+1.8}_{-0.6}$ & $0.8^{+1.0}_{-0.2}$ & $\mathbf{0.5}^{+0.7}_{-0.1}$ \\
    sine\_gaussian & $1.4^{+1.8}_{-0.6}$ & $8.9^{+13.5}_{-3.3}$ & $1.6^{+2.0}_{-0.7}$ & $40.2^{+16.3}_{-11.3}$ & $1.7^{+1.8}_{-0.8}$ & $0.6^{+0.7}_{-0.2}$ & $\mathbf{0.4}^{+0.6}_{-0.1}$ \\
    gaussian\_pulse & $\mathbf{0.4}^{+0.4}_{-0.2}$ & $3.8^{+3.8}_{-1.7}$ & $0.6^{+0.7}_{-0.3}$ & $37.7^{+13.9}_{-11.5}$ & $1.8^{+2.1}_{-0.8}$ & $0.6^{+0.7}_{-0.2}$ & $\mathbf{0.4}^{+0.4}_{-0.2}$ \\
    ringdown & $\mathbf{0.6}^{+0.8}_{-0.2}$ & $5.2^{+4.3}_{-2.1}$ & $0.9^{+1.0}_{-0.4}$ & $39.3^{+15.0}_{-9.5}$ & $2.2^{+2.4}_{-1.0}$ & $0.9^{+0.8}_{-0.4}$ & $0.7^{+0.6}_{-0.3}$ \\
    gengli\_H1 & $1.0^{+1.0}_{-0.3}$ & $3.8^{+3.3}_{-1.4}$ & $1.2^{+1.1}_{-0.4}$ & $36.6^{+13.0}_{-9.0}$ & $2.7^{+2.0}_{-1.1}$ & $1.2^{+0.8}_{-0.4}$ & $\mathbf{0.9}^{+0.7}_{-0.3}$ \\
    gengli\_L1 & $1.2^{+0.7}_{-0.4}$ & $4.0^{+2.8}_{-1.1}$ & $1.3^{+0.9}_{-0.4}$ & $36.1^{+10.9}_{-7.6}$ & $2.8^{+2.1}_{-1.1}$ & $1.2^{+0.9}_{-0.3}$ & $\mathbf{1.1}^{+0.7}_{-0.4}$ \\
    cdvgan\_blip & $1.2^{+1.0}_{-0.4}$ & $4.5^{+3.9}_{-1.7}$ & $1.4^{+1.1}_{-0.5}$ & $36.2^{+15.6}_{-9.1}$ & $2.9^{+2.6}_{-1.0}$ & $1.2^{+0.9}_{-0.4}$ & $\mathbf{0.9}^{+0.8}_{-0.2}$ \\
    cdvgan\_tomte & $1.0^{+0.7}_{-0.3}$ & $4.5^{+3.4}_{-1.7}$ & $1.2^{+0.8}_{-0.4}$ & $38.0^{+12.4}_{-10.7}$ & $2.6^{+2.5}_{-1.0}$ & $1.0^{+0.7}_{-0.3}$ & $\mathbf{0.9}^{+0.6}_{-0.3}$ \\
    cdvgan\_bbh & $1.0^{+1.1}_{-0.3}$ & $4.6^{+5.2}_{-1.5}$ & $1.1^{+1.2}_{-0.3}$ & $34.6^{+17.7}_{-6.7}$ & $2.7^{+2.4}_{-0.9}$ & $1.2^{+1.0}_{-0.5}$ & $\mathbf{0.8}^{+0.9}_{-0.2}$ \\
    cdvgan\_simplex & $1.8^{+1.4}_{-0.6}$ & $5.2^{+5.4}_{-1.6}$ & $2.0^{+1.6}_{-0.7}$ & $36.9^{+16.1}_{-7.6}$ & $3.5^{+3.5}_{-1.1}$ & $1.8^{+1.4}_{-0.5}$ & $\mathbf{1.6}^{+1.2}_{-0.5}$ \\
    cdvgan\_uniform & $1.6^{+1.1}_{-0.5}$ & $5.3^{+3.9}_{-1.9}$ & $1.7^{+1.3}_{-0.5}$ & $38.1^{+13.3}_{-10.0}$ & $3.1^{+3.3}_{-1.1}$ & $1.6^{+1.0}_{-0.5}$ & $\mathbf{1.4}^{+0.9}_{-0.5}$ \\
    \midrule
    \textbf{Total} & $1.2^{+1.1}_{-0.4}$ & $5.9^{+6.9}_{-2.4}$ & $1.4^{+1.3}_{-0.5}$ & $37.9^{+15.2}_{-9.2}$ & $2.5^{+2.3}_{-0.9}$ & $1.1^{+0.9}_{-0.4}$ & $\mathbf{0.9}^{+0.7}_{-0.3}$ \\
    \bottomrule
\end{tabular}
}
\caption{Median mismatch (\%) between injected and extracted glitches over 512 samples from each class. The bounds represent the range of the $1\sigma$ credible interval. Bold text highlights the best model for each signal type.}
\label{tab:match_table}
\end{table*}

\subsubsection{\label{sec:GravitySpy_experiments}Gravity Spy Glitches}
This experiment evaluates the performance of our model on \textit{Gravity Spy} glitches. 
We conduct a qualitative assessment of \textit{Gravity Spy} Q-scans before and after glitch subtraction to visually inspect the effectiveness of our model, while also inspecting the resulting glitch reconstruction in the time domain.
Since our fine-tuned \textit{DeepExtractor} was trained on $2\,$s samples extracted from $14\,$s whitened segments, we need to use $14\,$s of real data surrounding the glitch during inference to ensure the whitening filter remains consistent (see Section \ref{sec:real_backgrounds}). 
To avoid inadvertently suppressing the glitch during whitening, we estimate the PSD used for whitening from an adjacent $14\,$s segment via Welch's method that does not include the glitch. 
This approach ensures the glitch does not contaminate the PSD estimate, while still enabling effective whitening of the data.
Although the adjacent $14\,$s segment is not vetoed for other glitches (which could result in imperfect whitening), this approach helps prevent the desired glitch from being suppressed during the whitening process.

\subsubsection{\label{sec:gw_experiments}Reconstructing O3 signals}
To showcase the potential of \textit{DeepExtractor} for reconstructing real gravitational wave signals in a model-agnostic way, we apply it to three events from the O3 observing run: \textit{GW190521\_074359}, \textit{GW200129\_065458}, and \textit{GW200224\_222234}. 
The component masses for these events were, respectively, 45.4~$M_{\odot}$ and 33.4~$M_{\odot}$, 34.5~$M_{\odot}$ and 29.0~$M_{\odot}$, and 40.0~$M_{\odot}$ and 32.7~$M_{\odot}$.
These are among the loudest events reported during O3 that were confidently detected by both the Hanford and Livingston detectors.
We follow the same preprocessing pipeline as in the previous section: whitening a $14\,$s segment centered around the event, and extracting a $2\,$s window around the merger, with the merger time aligned at $1.85\,$s within the segment.
Similarly, we calculate the PSD from the previous $14\,$s segment to whiten the data, to ensure that the signal itself is not suppressed during the whitening process.
We then compare \textit{DeepExtractor}'s reconstructions to the maximum likelihood templates obtained from parameter estimation, once again using the time-domain mismatch as the evaluation metric.
The template itself is whitened using the same PSD as above, since the original template is given according to coloured detector noise.

\section{\label{sec:Results}Results}

\subsection{\label{sec:Simulated_experiments_results}Simulated Experiments}

\begin{figure*}[t] 
    \centering
    \includegraphics[width=\textwidth]{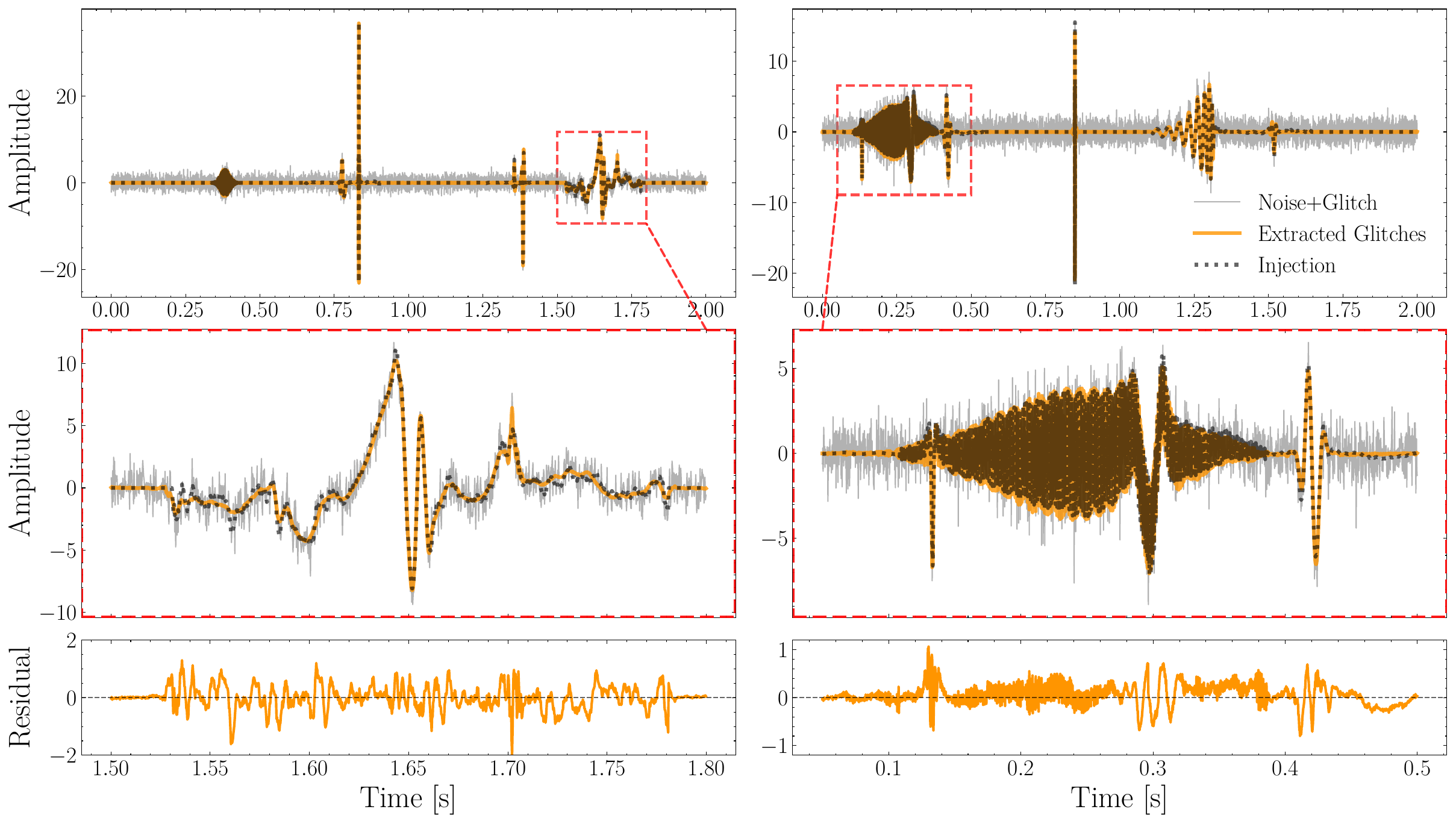}
    \caption{\textit{DeepExtractor} reconstructions (orange) of samples comprising multiple injections (black). The upper panels display the full $2\,$s of whitened strain (gray), the middle panels provide a zoomed-in view of the region outlined in red, emphasizing key features of the reconstructions and the bottom panels display the residual between the reconstructed and injected glitches.}
    \label{fig:signals_3}
\end{figure*}

Table \ref{tab:match_table} presents the median mismatch scores for each deep learning model across our test dataset, which consists of 512 samples from each class (totaling 6,144 test samples).
The bounds represent the $1\sigma$ credible interval.
All models in the table adhere to the framework outlined in FIG. \ref{fig:framework_overview}, where the background noise, $n(t)$, is first predicted and then subtracted from the input to reconstruct the glitch.
The mismatch distributions are generally skewed towards low values for most models.
However, a small subset of the samples includes very low SNR injections, located at the lower end of the SNR range (7.5).
These low-SNR injections subtly affect the background noise, resulting in significantly higher mismatch results for these specific cases across all models.
This trend is evident in the upper bounds shown in Table \ref{tab:match_table}, which are consistently larger than the corresponding lower bounds.

The high resolution ($257\text{x}257$) \textit{DeepExtractor} outperforms the other deep learning benchmarks with a median mismatch score of 0.9\% over the entire dataset. This marginally surpasses the ($129\text{x}129$) \textit{DeepExtractor}, which yields a median mismatch of 1.1\%, while the ($65\text{x}129$) \textit{DeepExtractor} yields a median mismatch of 2.5\%. 
These results suggest that higher-resolution magnitude and phase spectrograms lead to better generalization. 
While using even higher resolutions may yield further gains, the diminishing improvements observed across increasing resolutions indicate that returns may taper off. 
When mapping directly to the glitch target $g(t, f)$ using the ($129\text{x}129$) architecture, we yield an overall median mismatch of $1.7^{+1.1}_{-0.6}\%$, compared to the median mismatch of $1.1^{+0.9}_{-0.4}\%$ by first mapping to the background, $n(t, f)$. 
This indicates that our approach of mapping to the background noise and subsequently subtracting it yields better reconstruction performance compared to mapping directly to the simulated glitch space.

Although slightly trailing \textit{DeepExtractor} in performance, the time-domain variant, UNET1D, still delivers impressive results, achieving a median mismatch of 1.2\%. 
Similarly to \textit{DeepExtractor}, mapping directly to $g(t)$ with UNET1D does not improve the performance of mapping to $n(t)$, where a median mismatch of $1.3^{+1.2}_{-0.5}\%$ is obtained across the test dataset. 
This highlights the utility of our framework across time and spectrogram domains.
Furthermore, using both the dual-channel output and difference layer configuration make no notable improvement, with both yielding a median mismatch of 1.3\%.

In contrast to the Autoencoder1D model, which produces mismatch results similar to UNET1D, the Autoencoder2D model exhibits notably large mismatch values between the reconstructed and injected glitches. 
This suggests that feature compression within the Autoencoder layers—particularly in the spectrogram domain—may be a critical issue. 
To explore this further, we removed one of the Autoencoder2D layers to assess whether reduced compression could lead to better results in the spectrogram domain, but this adjustment did not improve the performance. 
This finding highlights the importance of skip connections in U-Nets, which play a vital role in preserving key features and enhancing reconstruction accuracy.

Finally, applying \textit{DeepExtractor} to 512 pure background samples without injections (i.e. where it is expected to reproduce the input) yields a maximum mismatch of $0.26\%$ between the target and the reconstructed background noise. 
This demonstrates the model's ability to accurately replicate the background noise in the absence of signals or glitches, ensuring that no extraneous signals or glitches are reconstructed when they are not present in the data.

FIG. \ref{fig:signals_3} shows reconstructions of multi-injection samples, similar to the training data, demonstrating \textit{DeepExtractor}'s versatility. Across 512 of such samples, \textit{DeepExtractor} achieves a median mismatch of $0.7^{+0.2}_{-0.2}$. The model excels with multi-injection samples by effectively matching non-Gaussian components, which are more frequent per sample and better aligned with the training data than the single-class samples in Table \ref{tab:match_table}\footnote{\textit{BayesWave} was not applied to these samples, as it requires a single trigger time, while these samples have multiple trigger times. We instead maintain a controlled comparison with single-class samples, where results are easier to interpret.}.

\subsection{\label{sec:BayesWave_comparison_results}BayesWave Comparison}

\begin{figure}[H]
\centering
\includegraphics[width =\linewidth]{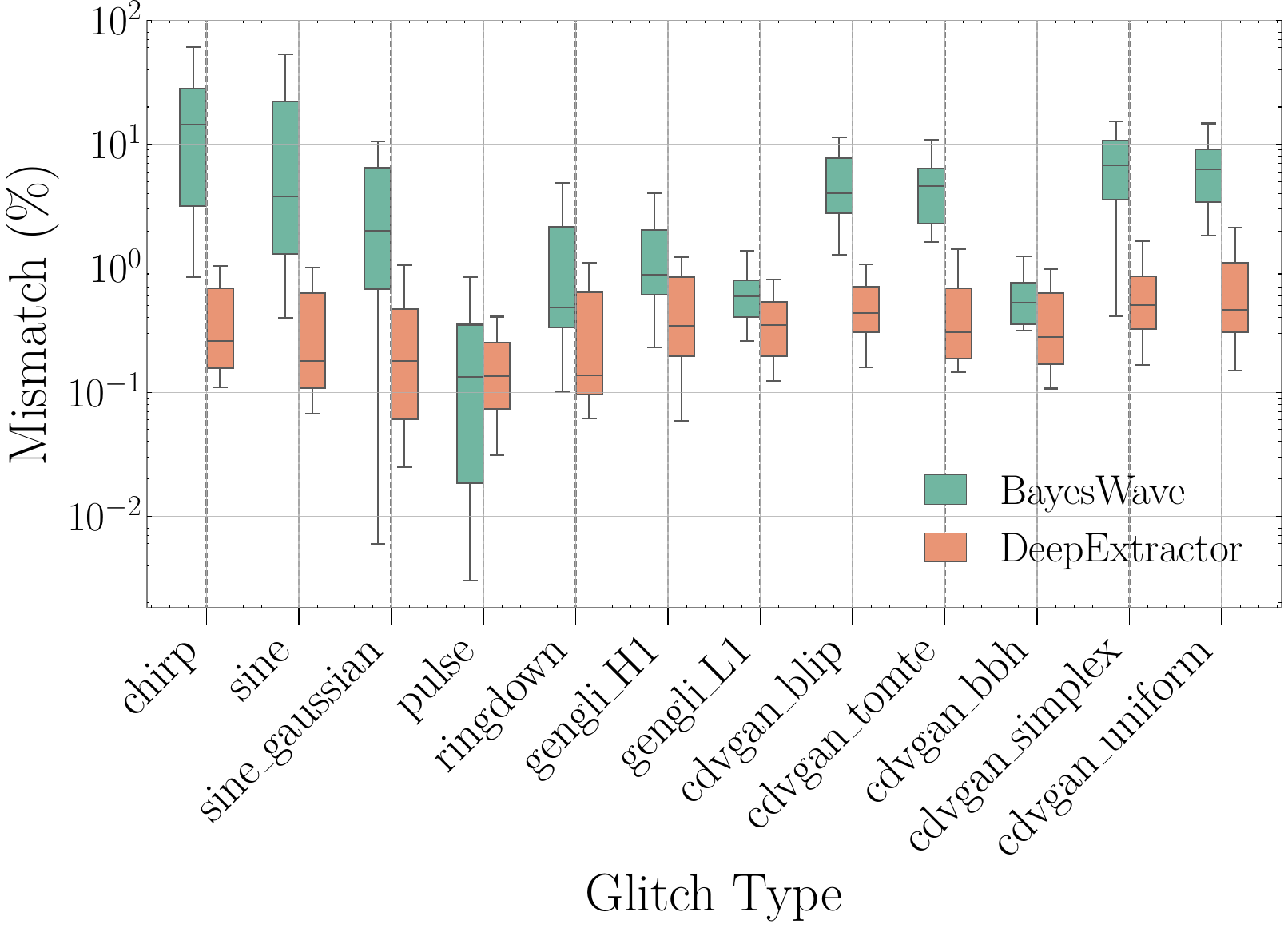}
\caption{Box plots showing mismatch between injected and reconstructed samples per class. Lower boxplots for \textit{DeepExtractor} indicate superior glitch reconstruction performance.}
\label{fig:BW_comparison_dist}
\end{figure}

\begin{figure*} 
    \centering
    \begin{subfigure}{0.32\linewidth}
        \centering
        \includegraphics[width=\linewidth]{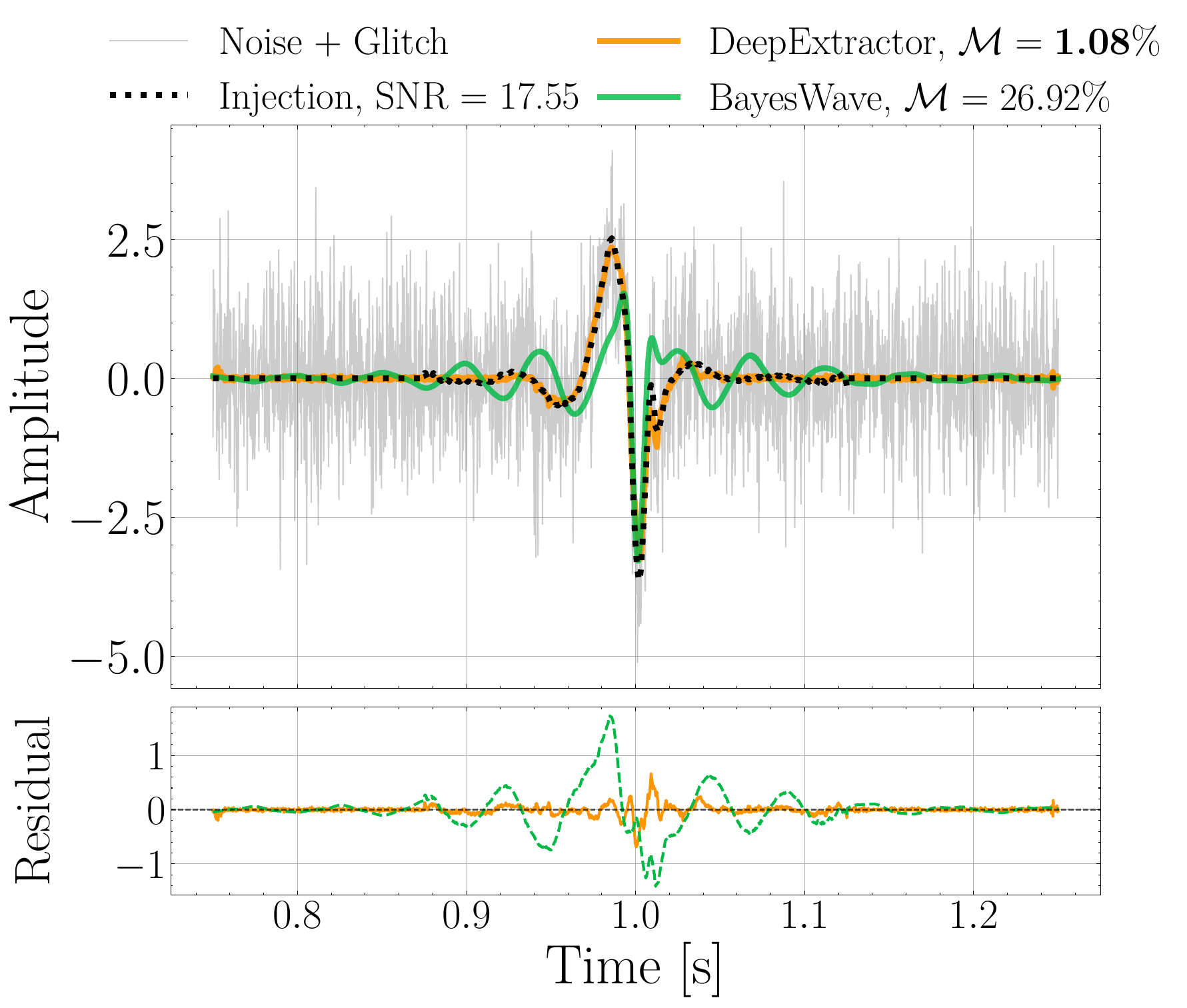}
        \label{fig:BW_output_1}
    \end{subfigure}
    \begin{subfigure}{0.32\linewidth}
        \centering
        \includegraphics[width=\linewidth]{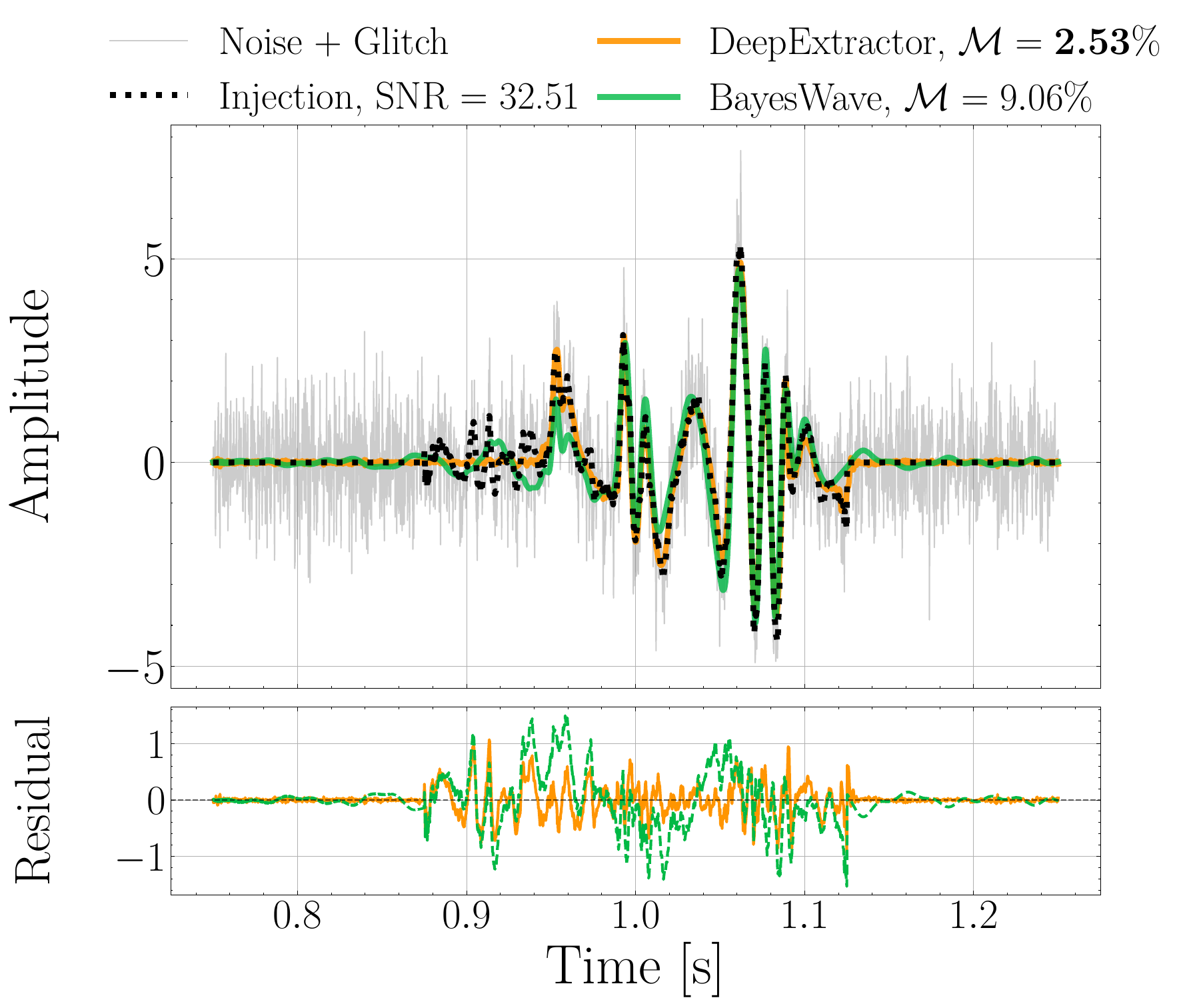}
        \label{fig:BW_output_2}
    \end{subfigure}
    \begin{subfigure}{0.32\linewidth}
        \centering
        \includegraphics[width=\linewidth]{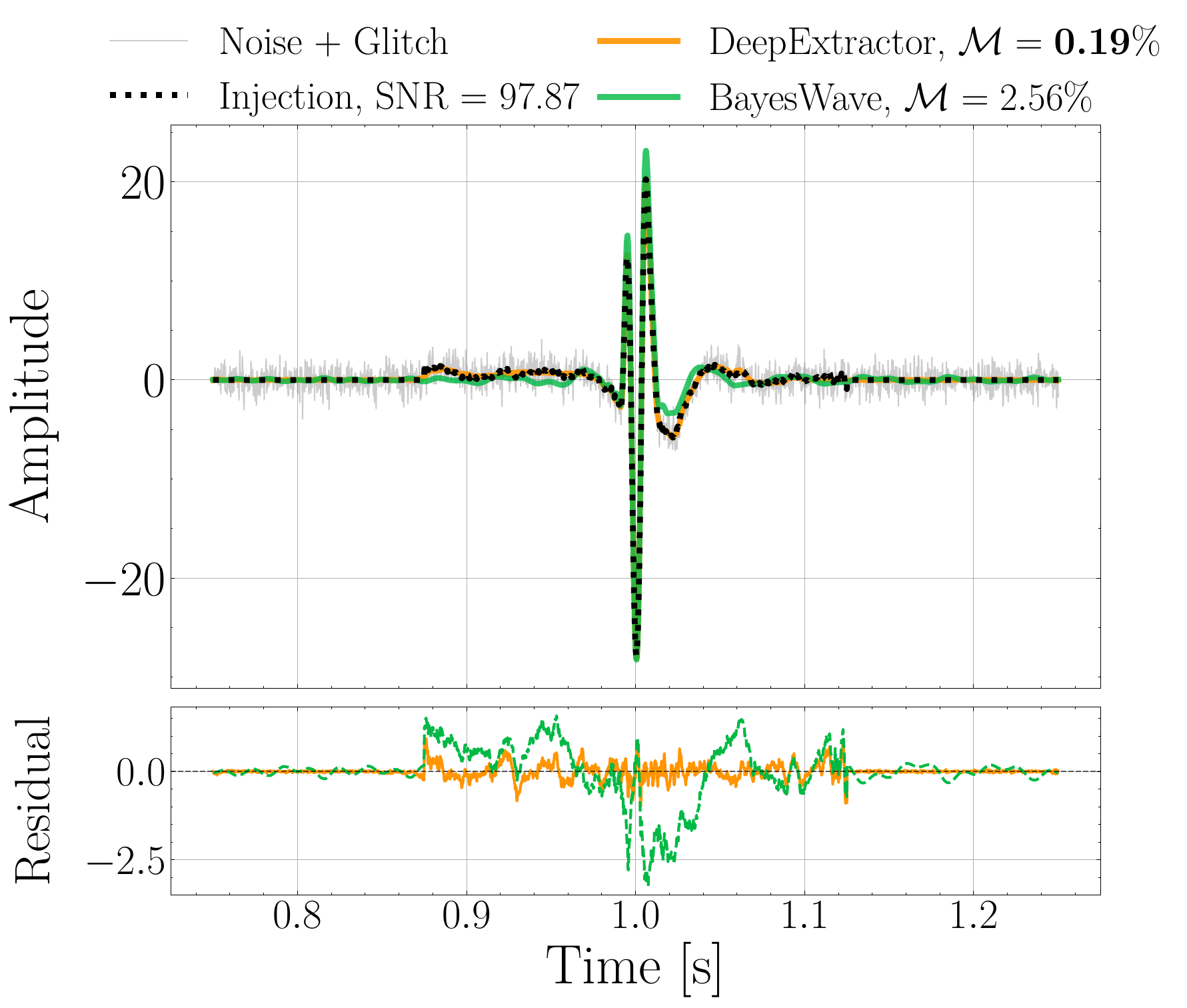}
        \label{fig:BW_output_3}
    \end{subfigure}
    \caption{Examples from our comparison between \textit{BayesWave} and \textit{DeepExtractor}. Injected SNRs and mismatches yielded from both approaches are shown above each plot. The bottom panels display the residual between the reconstructed and injected glitches for \textit{BayesWave} and \textit{DeepExtractor}.}
    \label{fig:BW_comparison}
\end{figure*}

FIG. \ref{fig:BW_comparison_dist} presents a comparison of mismatch performance between \textit{DeepExtractor} and \textit{BayesWave} on a reduced dataset, consisting of 30 samples per class. Examples of reconstructions from both \textit{BayesWave} and \textit{DeepExtractor} are shown in FIG. \ref{fig:BW_comparison}. \textit{DeepExtractor} consistently outperforms \textit{BayesWave}, even on previously unseen test classes. Specifically, \textit{DeepExtractor} achieves a median mismatch of $0.3^{+0.2}_{-0.1}\%$ across the entire test dataset, showing a significant improvement over \textit{BayesWave}, which has a median mismatch of $1.9^{+2.2}_{-1.1}\%$ (the bounds represent the $1\sigma$ confidence interval).
The performance improvement of \textit{DeepExtractor} compared to the previous section can be attributed to the lower SNR limit being raised to 15, making the reconstruction task easier.
 
The most notable improvements are observed in the \textit{chirp}, \textit{sine}, \textit{sine-gaussian}, and \textit{ringdown} classes, where \textit{BayesWave} reaches maximum mismatches as high as $81.8\%$, $66.8\%$, $72.0\%$, and $83.9\%$, respectively. 
In contrast, \textit{DeepExtractor} shows much lower maximum mismatches of $4.0\%$, $3.0\%$, $1.9\%$, and $16.5\%$ for these classes.

The discrepancies may be due to \textit{BayesWave}'s difficulty in converging due to high-frequency features in these samples. Furthermore, \textit{DeepExtractor} demonstrates significant improvements when applied to realistic glitch samples from glitch generators, particularly hybrid samples from \textit{cDVGAN}, which feature diverse glitch morphologies.

Finally, \textit{DeepExtractor} also offers a substantial advantage in terms of speed, reconstructing a single glitch sample in an average of $0.14$ seconds on a CPU (or $0.05$ seconds per glitch when processing a batch of 256 samples). 
This represents a speedup of over $10,000$ times compared to \textit{BayesWave}. 

The significantly longer runtime of \textit{BayesWave} stems from its use of RJMCMC to explore a multidimensional parameter space and to construct posterior distributionss.
While computationally intensive, this Bayesian approach provides uncertainty estimates and flexible modeling of both glitches and astrophysical signals with minimal assumptions.

In contrast, \textit{DeepExtractor} performs fast, deterministic inference via a single forward pass through a trained neural network, bypassing the need for iterative sampling or likelihood evaluation. 
This makes it well-suited for real-time or large-scale applications.
While it trades some flexibility and uncertainty quantification for speed, \textit{DeepExtractor} offers rapid, high-fidelity reconstructions of glitches in scenarios where fast turnaround is critical or full Bayesian inference is computationally prohibitive.

\subsection{\label{sec:GravitySpy_experiments_results}Gravity Spy Glitches}

\begin{figure*}
\centering
\begin{subfigure}[t]{0.24\linewidth}
  \centering
  \includegraphics[width = \linewidth]{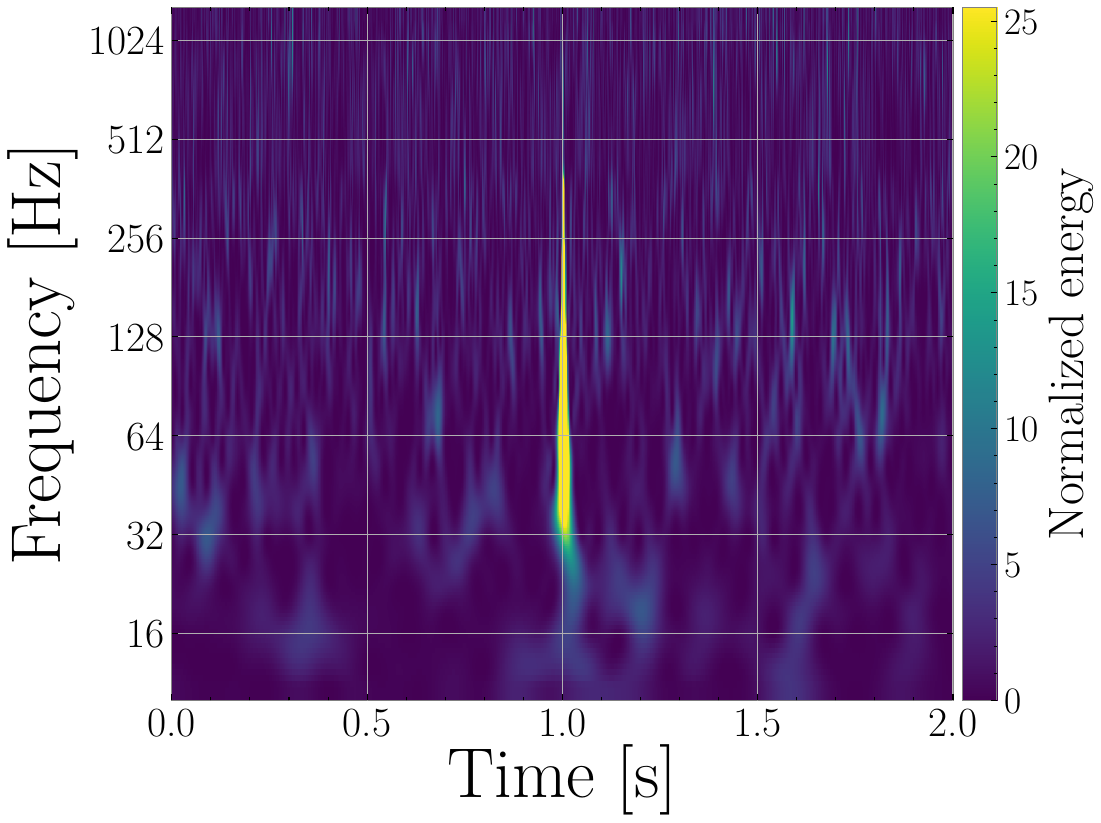}
  \caption{Input Glitch}
  \label{fig:gspy_input}
\end{subfigure}%
\begin{subfigure}[t]{0.24\linewidth}
  \centering
  \includegraphics[width = \linewidth]{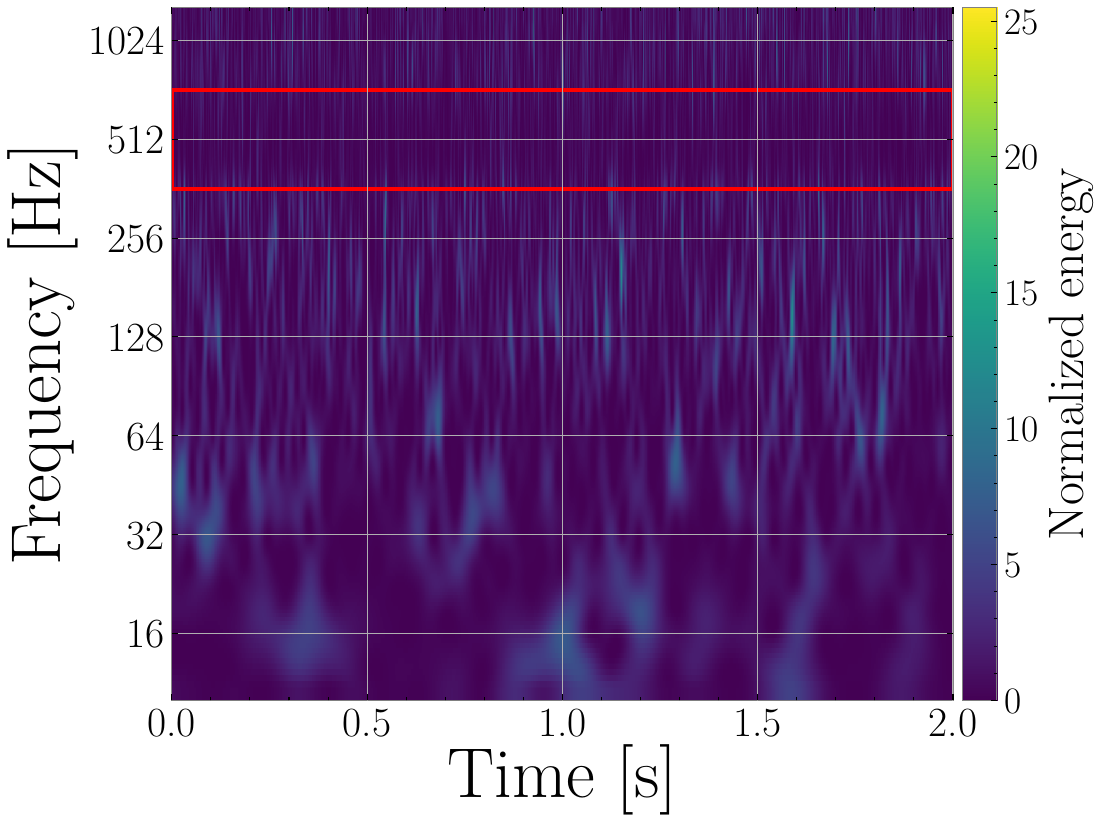}
  \caption{Without transfer learning}
  \label{fig:gspy_output_no_tl}
\end{subfigure}%
\begin{subfigure}[t]{0.24\linewidth}
  \centering
  \includegraphics[width = \linewidth]{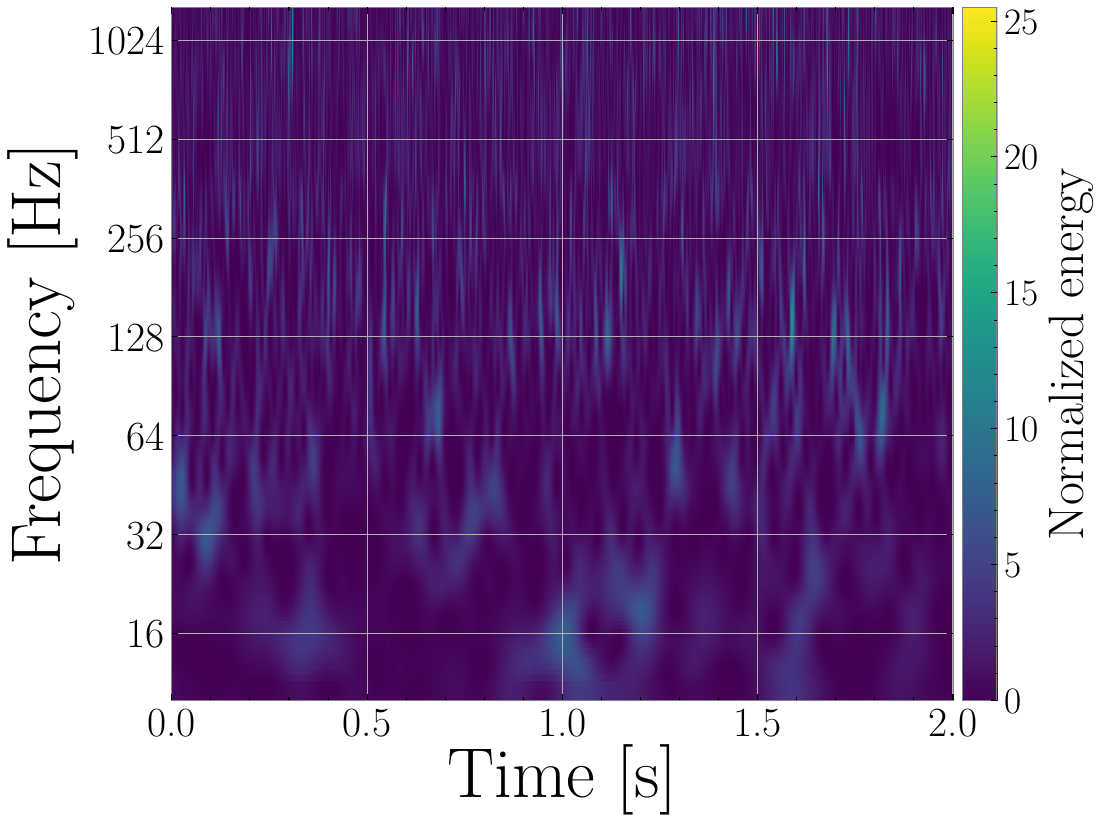}
  \caption{With transfer learning}
  \label{fig:gspy_output}
\end{subfigure}%
\begin{subfigure}[t]{0.24\linewidth}
  \centering
  \includegraphics[width = \linewidth]{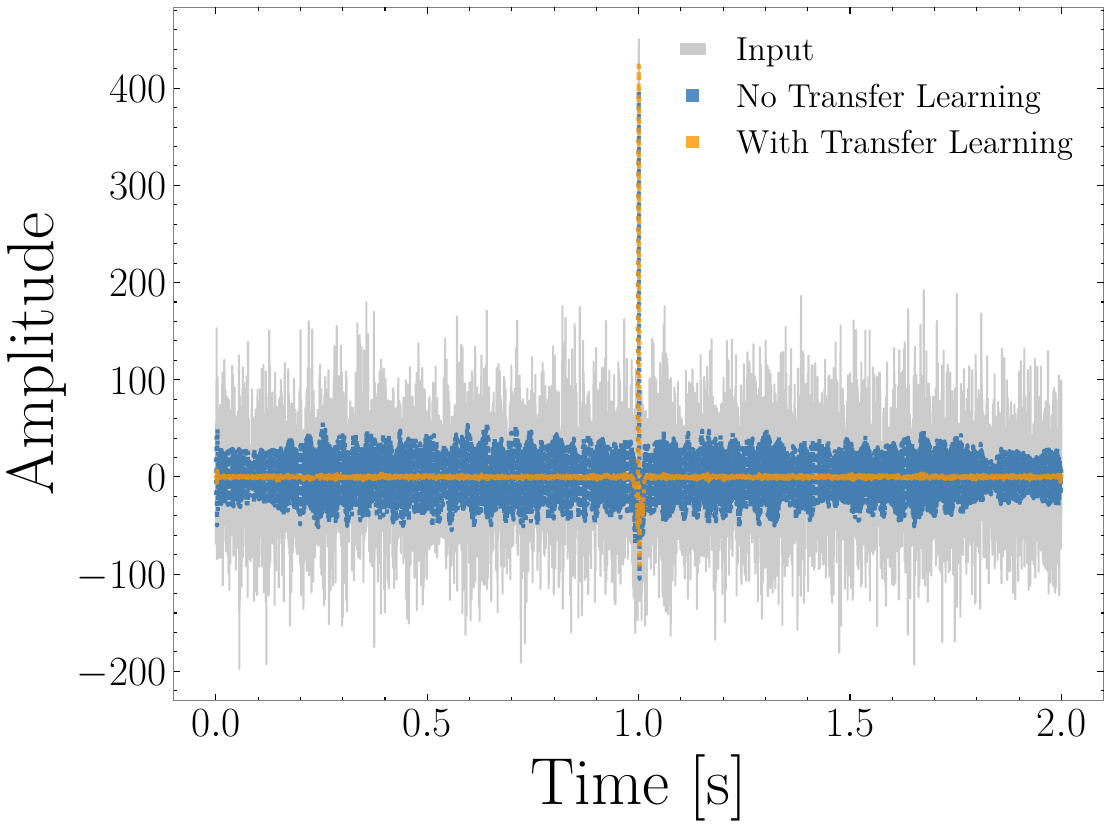}
  \caption{Time-series outputs}
  \label{fig:gspy_ts}
\end{subfigure}%
\caption{An example of subtracting a blip glitch with and without transfer learning \textit{DeepExtractor} on real detector noise. FIG. \ref{fig:gspy_output_no_tl} shows that, without transfer learning on real detector noise, the line features at $512\,$Hz are also subtracted, highlighted within the red lines. This is because the model was trained on a purely flat, simulated PSD, causing the realistic line features to also be reconstructed as part of the excess power.}
\label{fig:gspy_transfer_learning}
\end{figure*}


\begin{figure*}
\centering
\includegraphics[width=0.95\textwidth]{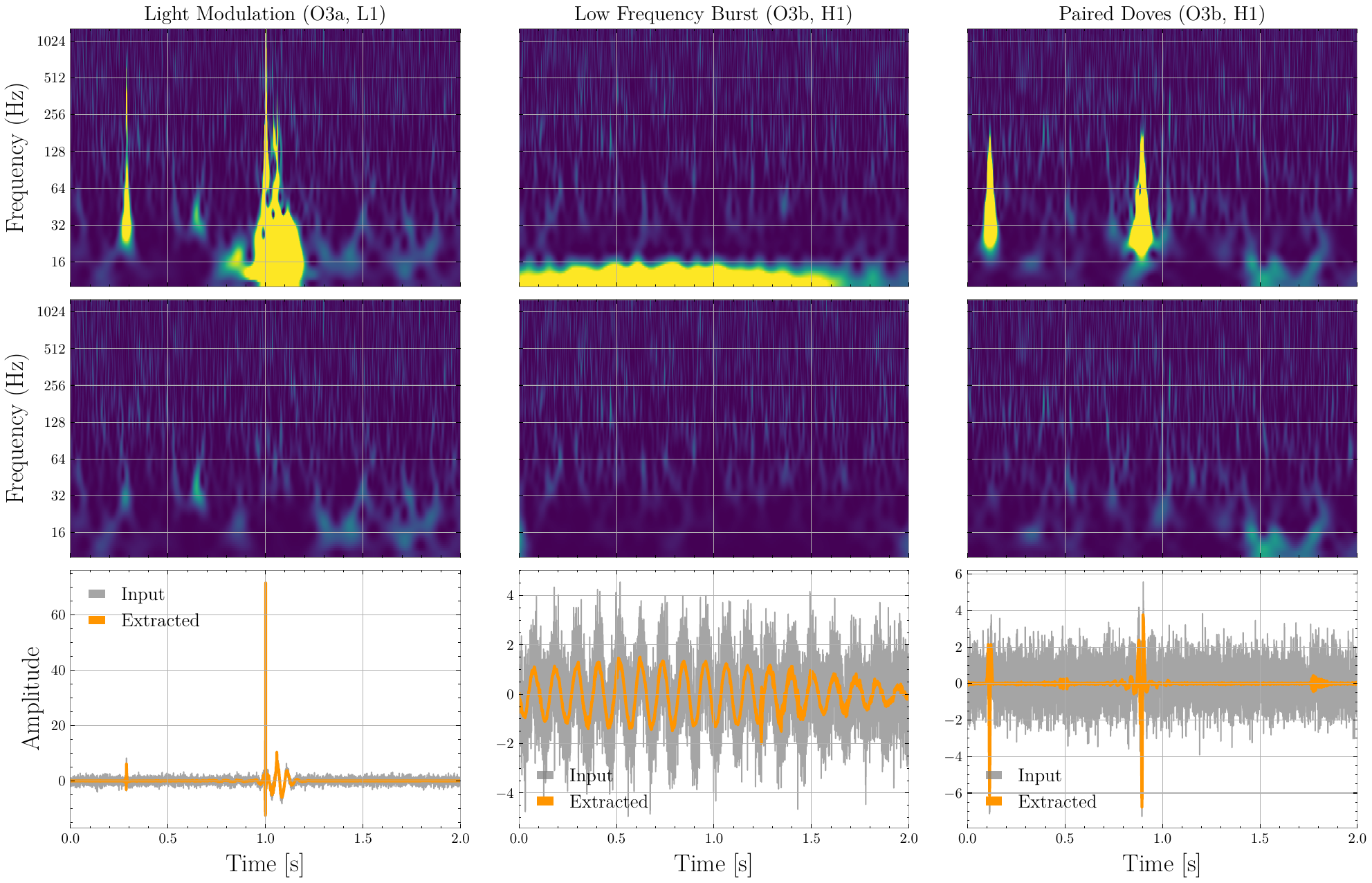}
\caption{Reconstructions for three examples from distinct \textit{Gravity Spy} glitch classes: \textit{Light Modulation}, \textit{Low Frequency Burst}, and \textit{Paired Doves}. The top row displays Q-scans of the input to the network. The middle row shows Q-scans of the residual after subtracting \textit{DeepExtractor}’s reconstruction. The bottom row presents the corresponding time-domain input and reconstructed waveforms. The maximum color limit in the Q-scans is set to 25, similarly to the Q-scans shown in FIG. \ref{fig:gspy_transfer_learning}.}
\label{fig:gspy_examples_3}
\end{figure*}

This experiment applies transfer learning to fine-tune \textit{DeepExtractor} on real LIGO O3 data from both the Hanford and Livingston detectors. The primary objective is to improve the model’s ability to reconstruct real glitch events. Given the complex, unmodeled nature of glitches, we take a qualitative approach by visualizing \textit{DeepExtractor's} reconstruction performance through Q-scans and time-series plots.

FIG. \ref{fig:gspy_transfer_learning} illustrates the reconstruction of a blip glitch from LIGO Hanford during the O3a run, before and after transfer learning with \textit{DeepExtractor}. This example highlights the significance of further training with real detector data, as it includes features such as high-frequency lines that are common in actual LIGO data. Incorporating these realistic features is crucial for the model to accurately differentiate between detector noise and genuine glitch features, improving its reconstruction capabilities.

FIG. \ref{fig:gspy_transfer_learning}d displays the time-domain output of both the pre-trained and fine-tuned models. Without transfer learning, the model erroneously reconstructs the line features, obscuring the true glitch signal. However, after transfer learning, the model compensates for these features, resulting in a more accurate glitch reconstruction.

FIG \ref{fig:gspy_examples_3} presents examples of \textit{DeepExtractor} reconstructions following transfer learning, illustrating performance across three distinct glitch classes. Additional reconstruction examples are provided in Appendix \ref{sec:gspy_glitch_examples}.
The Q-scans show that, for most glitch classes, noise features are effectively preserved after glitch removal. However, for louder glitches, such as those from the `Extremely Loud' and `Koi Fish' classes, where the glitch dominates the sample, the removal process leads to a `zeroing out' effect on the data, as seen in the Q-scans post-mitigation. However, in most cases, the time-series plots demonstrate that the reconstructions align closely with the actual data. Overall, \textit{DeepExtractor} exhibits flexibility across glitch classes.

\subsection{\label{sec:gw_results}Reconstructing O3 signals}
In this section, we present the reconstructions of three gravitational-wave events detected during O3 observing run: \textit{GW190521\_074359}, \textit{GW200129\_065458}, and \textit{GW200224\_222234}.
\textit{DeepExtractor} is applied independently to whitened strain data from each detector around the time of merger.

\begin{figure*}[t]
\centering

\begin{subfigure}{\linewidth}
    \centering
    \includegraphics[width=\linewidth]{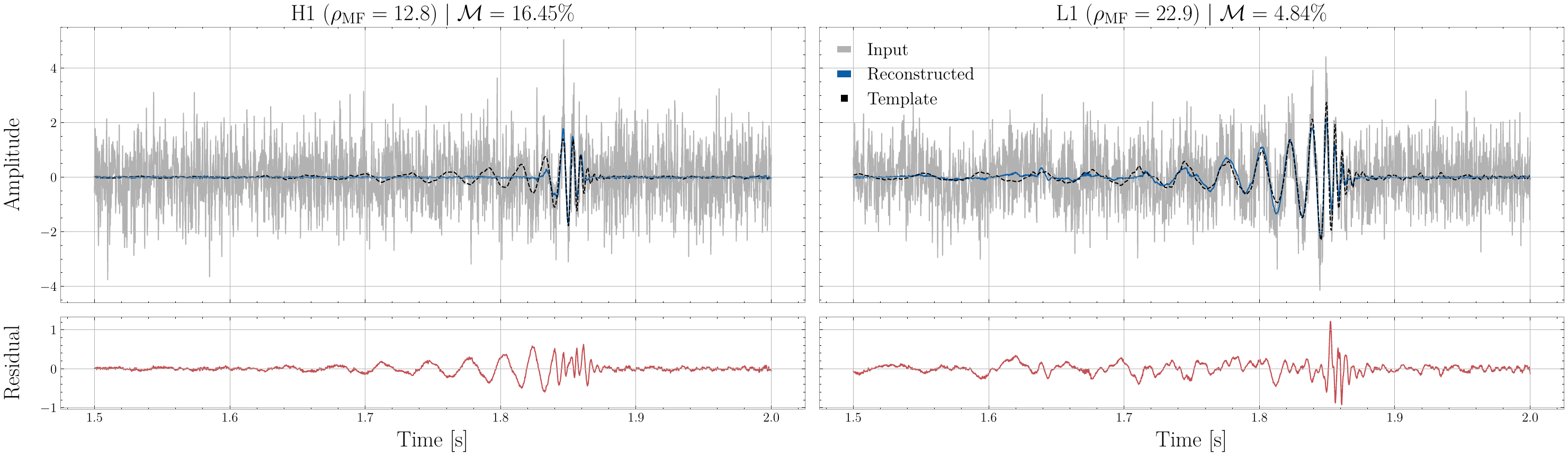}
    \caption{\textit{GW190521\_074359}}
    \label{fig:gw190521}
\end{subfigure}

\vspace{1em}

\begin{subfigure}{\linewidth}
    \centering
    \includegraphics[width=\linewidth]{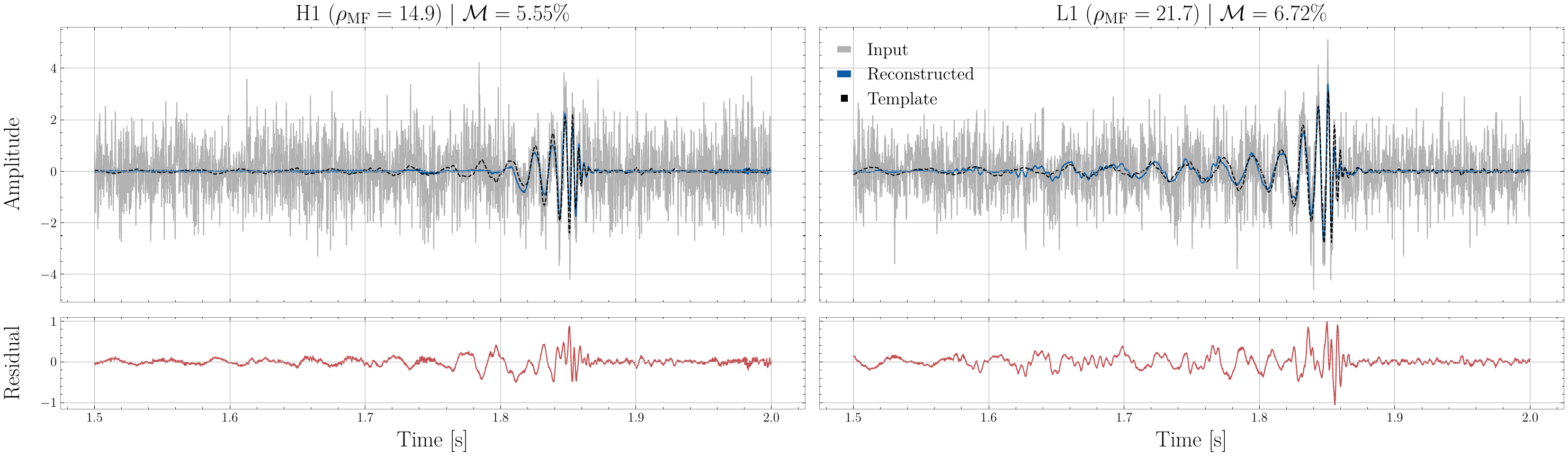}
    \caption{\textit{GW200129\_065458}}
    \label{fig:gw200129}
\end{subfigure}

\vspace{1em}

\begin{subfigure}{\linewidth}
    \centering
    \includegraphics[width=\linewidth]{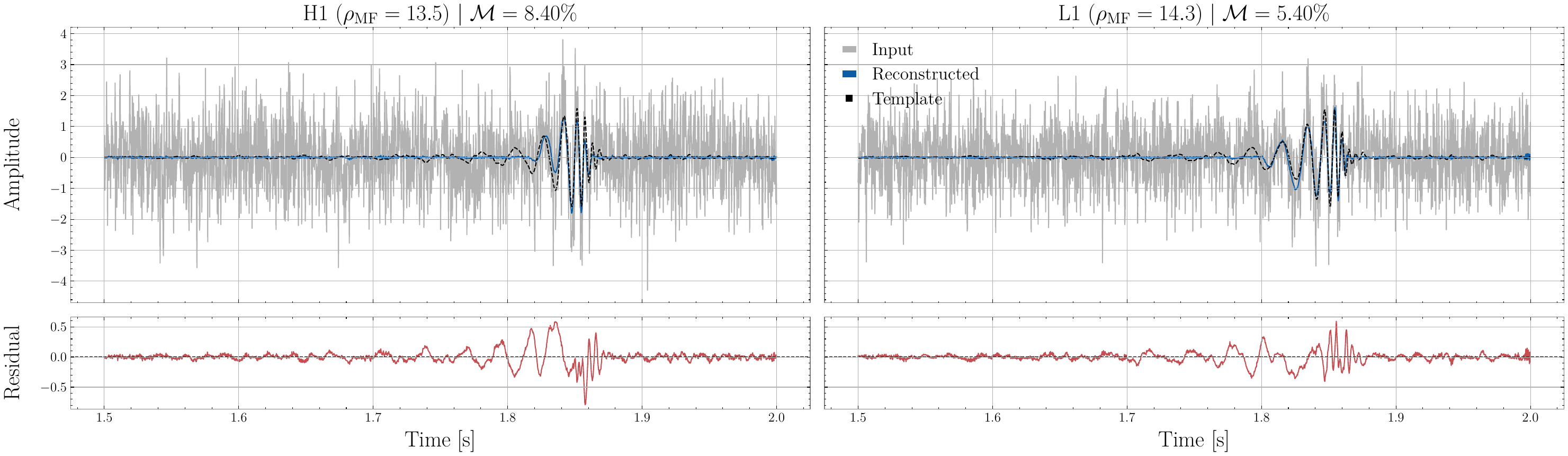}
    \caption{\textit{GW200224\_222234}}
    \label{fig:gw200224}
\end{subfigure}

\caption{Reconstruction of gravitational-wave signals from three O3 events using \textit{DeepExtractor}. 
Each panel shows the comparison between the reconstructed waveform, the template, and the residuals for H1 and L1 detectors. 
The matched-filter SNR ($\rho_{MF}$) in each detector frame is shown at the top of the respective plots, along with the mismatch ($\mathcal{M}$) between the reconstruction and the maximum likelihood template.
We acknowledge that misalignment between our data preprocessing and that of the parameter estimation pipeline likely contributes to the observed mismatches.}
\label{fig:GW_reco_all}
\end{figure*}

FIG.~\ref{fig:GW_reco_all} shows the reconstructed signals for all three events in both LIGO detectors. For each case, we report the SNR, the mismatch relative to the maximum likelihood waveform \cite{GWTC_3_data}, and the residual between the reconstructed and the maximum likelihood waveform.

Although \textit{DeepExtractor} was not explicitly trained on astrophysical signals and is agnostic to their morphologies, it achieves mismatches below 10\% for most cases. The exception is the \textit{GW190521\_074359} event in the Hanford detector, where the mismatch reaches 16.45\%.

This degradation in performance compared to the simulated setting is likely due to a misalignment between our preprocessing pipeline and that of the parameter estimation pipeline—specifically our whitening approach for both strain data and parameter estimation templates, which is necessary for input to \textit{DeepExtractor}.
Improving the alignment of \textit{DeepExtractor's} training scheme with parameter estimation pipelines and incorporating signal models into training could enhance the performance.

This experiment demonstrates the promise of \textit{DeepExtractor} as a model-agnostic tool for reconstructing gravitational-wave signals.
In cases where glitches are coincident with signals, \textit{DeepExtractor} reconstructs both the signal and glitch simultaneously—a scenario we also explore in Appendix \ref{sec:gw_glitch_appendix}.
Future work will involve extending \textit{DeepExtractor} to separate both GWs and glitches during reconstruction (see Section \ref{sec:Conclusions} for more discussion).

\section{\label{sec:Conclusions}Conclusions and Future Work}
This paper introduced \textit{DeepExtractor}, a novel deep learning framework designed to reconstruct signals and glitches in gravitational wave detector data. Leveraging a PSD-informed U-Net architecture, \textit{DeepExtractor} processes magnitude and phase spectrograms derived from the short-time Fourier transform (STFT). This spectrogram-based approach enables transformations between the frequency and time domains, essential for tasks such as glitch mitigation and signal reconstruction.

The key innovation lies in the training strategy. Rather than directly mapping input data to clean training waveforms, the model is trained to isolate the noise component from input data comprising background noise and injected signals or glitches. 
By subtracting the predicted noise, \textit{DeepExtractor} reconstructs the signal or glitch. This approach outperforms direct mapping in generalization. The framework is highly flexible, capable of reconstructing any power excess above the inherent detector noise. This flexibility stems from training on a diverse dataset of linear combinations of waveforms from five distinct analytical classes (proxy glitches), enabling effective interpolation to reconstruct unseen waveforms.

Four experiments validated \textit{DeepExtractor's} efficacy. The first experiment involves simulated Gaussian noise with injected glitches, measuring time-domain mismatch performance across several deep learning models. 
The results demonstrate that our U-Net architecture significantly outperforms autoencoders without skip connections, even when the overall architectures are otherwise identical. 
Additionally, U-Nets operating on magnitude and phase spectrograms are shown to deliver superior performance compared to a 1D U-Net applied directly to the corresponding time-series data.

In the second experiment, we compared \textit{DeepExtractor} to the state-of-the-art \textit{BayesWave} algorithm in a simulated environment. 
Our results reveal that \textit{DeepExtractor} not only delivers higher accuracy and greater stability than \textit{BayesWave} but also demonstrates  a key advantage of deep learning over traditional Bayesian approaches: computational efficiency. 
While \textit{BayesWave} relies on computationally intensive Bayesian inference, \textit{DeepExtractor} achieves a remarkable speedup of over 10,000 times on a CPU, demonstrating its potential for real-time gravitational wave data analysis.

We evaluated \textit{DeepExtractor's} capability to reconstruct real glitch events from LIGO's third observing run. 
This represents the first comprehensive application of deep learning to glitch reconstruction. 
By utilizing Q-scan and time-domain visualizations, we ensured effective glitch removal while preserving the underlying noise characteristics. 
The study highlighted the importance of integrating PSD knowledge through transfer learning, allowing the model to adapt effectively to the complexities of real detector data.

Finally, we applied \textit{DeepExtractor} to three real gravitational-wave events from LIGO’s third observing run, demonstrating its potential as a model-agnostic tool for signal reconstruction. 
The resulting reconstructions showed mismatches under 10\% against the most of the parameter estimation templates, even without being explicitly trained on GW signals.
Some of the observed mismatch is likely attributable to differences between our preprocessing pipeline and that used in parameter estimation. 
Aligning the preprocessing steps more closely with standard parameter estimation workflows, and incorporating signal models during training, may further enhance the reconstruction performance.

Several avenues remain open for future research. For example, modifying the training classes or tuning associated hyperparameters—such as the duration range of $[0.125, 2]\,$s—could help assess whether these adjustments yield improved reconstruction performance. 
While \textit{BayesWave} remains the most widely used tool for glitch mitigation in the gravitational-wave community, future work could compare \textit{DeepExtractor} with alternative approaches, such as \textit{gwsubtract}, specifically in cases where glitches exhibit linear correlations with auxiliary channels. 
Comparisons with other deep learning–based signal reconstruction methods, such as AWaRe, are also of interest. 
However, to be applicable for arbitrary glitch reconstruction, such models would need to be adapted to incorporate \textit{DeepExtractor}’s training scheme—or a comparable approach capable of recovering excess power directly from strain data.

For signal reconstruction, a key next step is to compare \textit{DeepExtractor's} performance against established search and parameter estimation pipelines. 
While the model can reconstruct both a signal and glitch when they overlap in the data, it does not yet support their separation. 
This limitation could be addressed by extending the training framework to separately reconstruct overlapping components. 
Rather than isolating the noise only, the model could be trained to also output glitch and signal estimates, using simulated signals from standard waveform models as supervised targets. Furthermore, incorporating data from multiple detectors would allow \textit{DeepExtractor} to model coherent signal power across the network, potentially improving reconstruction accuracy in cases where glitches and signals coincide.

Recent research leveraging the null stream in the Einstein Telescope’s triangular design has shown effective glitch mitigation in the presence of overlapping signals, marking a significant advancement for the third-generation detector era \cite{Harsh_null}. 
Building on this progress, \textit{DeepExtractor} could further enhance the null stream’s performance by precisely isolating uncorrelated glitches from signals. 
Integrating \textit{DeepExtractor} into this framework would be a straightforward process and could strongly reinforce the scientific case for the Einstein Telescope’s triangular design.

\section{Acknowledgements}
This research has made use of data or software obtained from the Gravitational Wave Open Science Center (gwosc.org), a service of the LIGO Scientific Collaboration, the Virgo Collaboration, and KAGRA. This material is based upon work supported by NSF's LIGO Laboratory which is a major facility fully funded by the National Science Foundation, as well as the Science and Technology Facilities Council (STFC) of the United Kingdom, the Max-Planck-Society (MPS), and the State of Niedersachsen/Germany for support of the construction of Advanced LIGO and construction and operation of the GEO600 detector. Additional support for Advanced LIGO was provided by the Australian Research Council. Virgo is funded, through the European Gravitational Observatory (EGO), by the French Centre National de Recherche Scientifique (CNRS), the Italian Istituto Nazionale di Fisica Nucleare (INFN) and the Dutch Nikhef, with contributions by institutions from Belgium, Germany, Greece, Hungary, Ireland, Japan, Monaco, Poland, Portugal, Spain. KAGRA is supported by Ministry of Education, Culture, Sports, Science and Technology (MEXT), Japan Society for the Promotion of Science (JSPS) in Japan; National Research Foundation (NRF) and Ministry of Science and ICT (MSIT) in Korea; Academia Sinica (AS) and National Science and Technology Council (NSTC) in Taiwan.
The authors are grateful for computational resources provided by the LIGO Laboratory and supported by the National Science Foundation Grants No. PHY-0757058 and No. PHY-0823459.
Finally, we thank the anonymous referees for their useful suggestions in improving the manuscript.

\clearpage

\onecolumngrid

\section{Appendixes}

\vspace{-3mm}

\subsection{\label{sec:simulated_glitch_descriptions}Simulated training glitches}

Table \ref{tab:functions} outlines the analytical models employed to generate the proxy glitch classes for training \textit{DeepExtractor}, along with the parameters used in their functions.

\begin{table}[h!]
\centering
\begin{tabular}{|l|p{7cm}|p{7cm}|}
\hline
\textbf{Function} & \textbf{Description} & \textbf{Key Parameters} \\ \hline
\textit{chirp} & Generates a linear chirp signal with frequency varying from \texttt{f0} to \texttt{f1} over the duration. & \texttt{f0\_min, f0\_max, f1\_min, f1\_max, duration, sample\_rate} \\ \hline
\textit{sine} & Generates a simple sine wave with a frequency randomly chosen between \texttt{freq\_min} and \texttt{freq\_max}. & \texttt{freq\_min, freq\_max, duration, sample\_rate} \\ \hline
\textit{sine-gaussian} & Generates a sine wave modulated by a Gaussian envelope. & \texttt{freq\_min, freq\_max, duration, sample\_rate} \\ \hline
\textit{Gaussian pulse} & Generates a Gaussian pulse with randomized frequency and bandwidth characteristics. & \texttt{fc\_min, fc\_max, bw\_min, bw\_max, bwr\_min, bwr\_max, tpr\_min, tpr\_max, duration, sample\_rate} \\ \hline
\textit{ringdown} & Generates a ringdown signal based on a damped sinusoidal model with a random frequency \texttt{f0}, decay time \texttt{tau}, and quality factor \texttt{Q}. There is also a 50\% probability to flip the signal horizontally. & \texttt{duration, sample\_rate, n\_signals} \\ \hline
\end{tabular}
\caption{Summary of analytical glitch models used to simulate glitches for training \textit{DeepExtractor}.}
\label{tab:functions}
\end{table}

\subsection{\label{sec:STFT_parameters}STFT Parameters}
Table \ref{tab:parameters} details the parameters and their respective values used for STFT computation, enabling transformations to and from the time-frequency (spectrogram) domain.

\begin{table}[h!]
\centering
\begin{tabular}{|l|p{7cm}|p{5cm}|}
\hline
\textbf{Parameter} & \textbf{Description} & \textbf{Value} \\ \hline
\texttt{n\_fft} & The size of the Fast Fourier Transform (FFT), typically a power of 2. This determines the resolution of the frequency analysis. & 512 \\ \hline
\texttt{window\_length} & The length of the window function used in the FFT. This determines how much of the signal is considered at once. & 64 \\ \hline
\texttt{hop\_length} & The step size between consecutive windows when performing the FFT. It defines how much overlap there is between consecutive windows. & 32 \\ \hline
\textbf{Hann window} & A window function often used in FFT analysis due to its smooth tapering, reducing spectral leakage. It satisfies the COLA condition when paired with a hop length of half the window length. & Used \\ \hline
\textbf{COLA condition} & The Constant Overlap-Add (COLA) condition ensures that the signal is reconstructed without any discontinuities, preserving the signal's continuity when the windowed segments are overlapped. & Compliance ensured through the combination of a Hann window and $\texttt{hop\_length} = \texttt{window\_length}/2$. \\ \hline
\end{tabular}
\caption{Overview of parameters used in the short-time Fourier transform (STFT) for transforming between the time and spectrogram domains in \textit{DeepExtractor}.}
\label{tab:parameters}
\end{table}

\newpage

\subsection{\label{sec:cDVGAN generations}Test samples generated by \textit{gengli} and \textit{cDVGAN}}
\begin{figure}[H]
\centering
\includegraphics[width =0.45\linewidth]{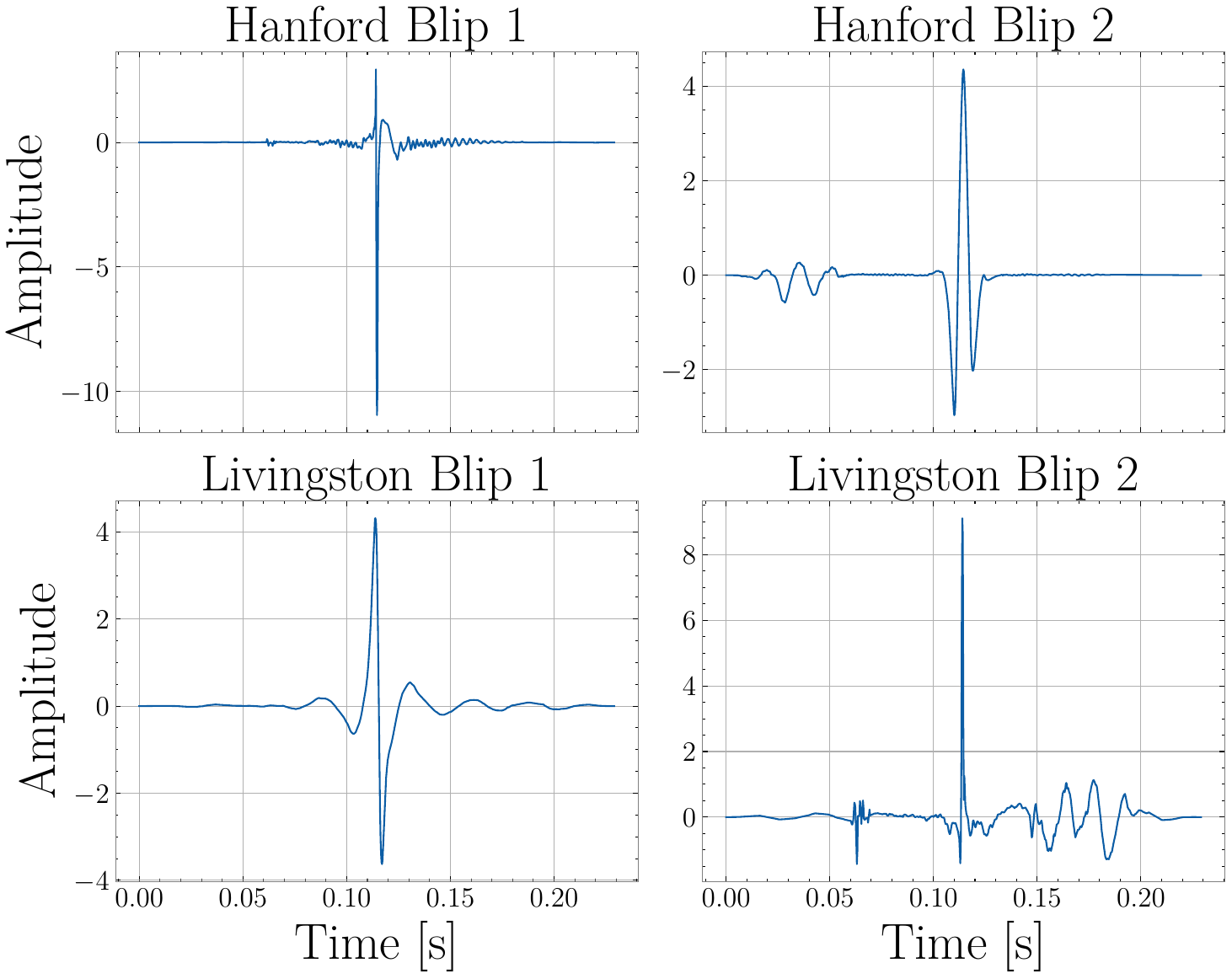}
\caption{Two example Blip glitches from the Hanford (top row) and Livingston (bottom row) detectors, generated using \textit{gengli}.}
\label{fig:gengli_examples}
\end{figure}

\begin{figure}[H]
\centering
\includegraphics[width =0.58\linewidth]{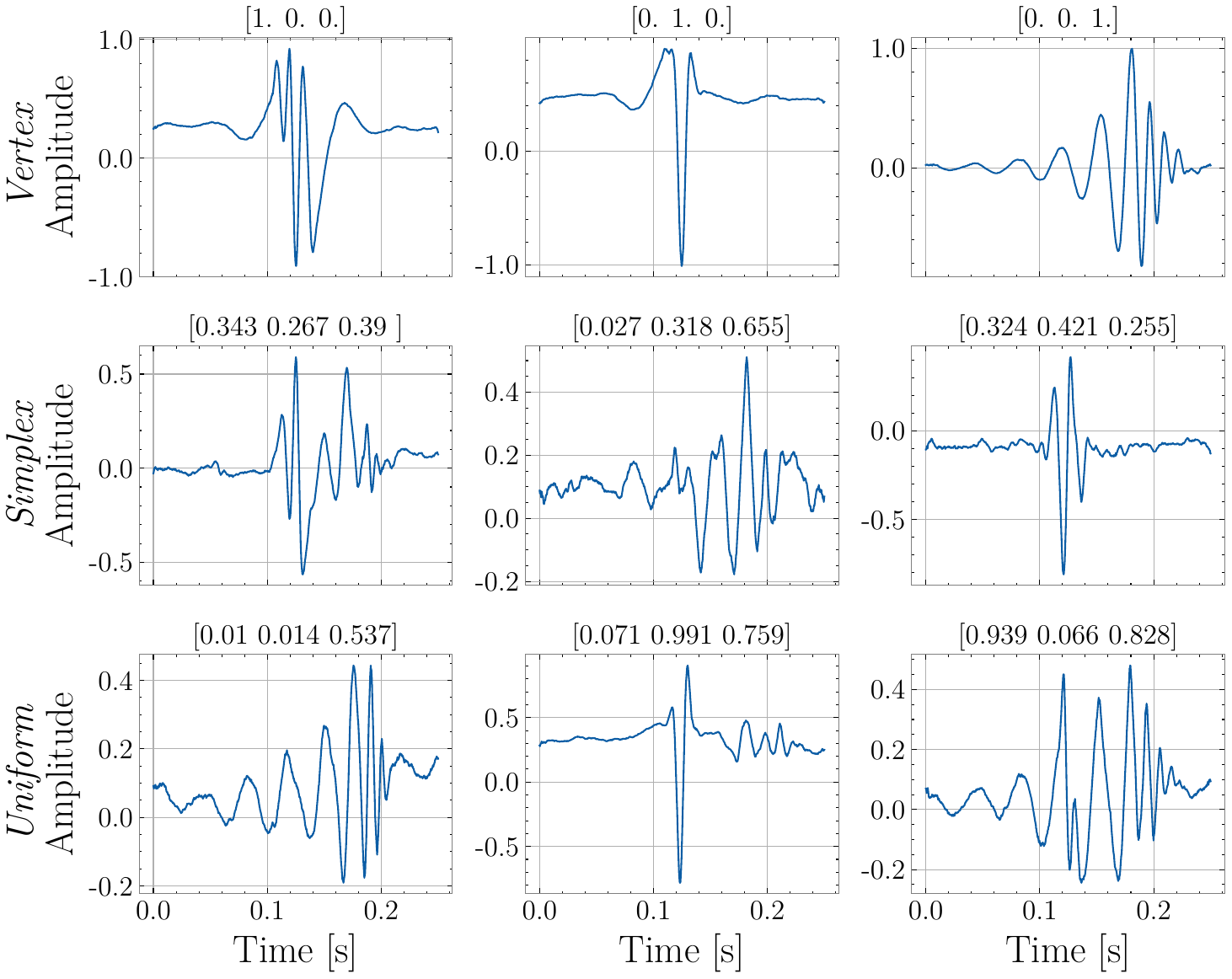}
\caption{Examples of in-class (vertex) and class-mixed samples (simplex, uniform) generated by cDVGAN. The class vector used to generate each sample is shown at the top of each plot.}
\label{fig:cdvgan_signals}
\end{figure}

FIG. \ref{fig:gengli_examples} and \ref{fig:cdvgan_signals} show samples generated by \textit{gengli} and \textit{cDVGAN} respectively, used to evaluate the generalization capability of \textit{DeepExtractor}. 
By interpolating across the class space, cDVGAN enables the creation of a diverse dataset that captures key features characteristic of both gravitational-wave signals and glitches. 
It is observed that by sampling the class vector, we can achieve a wide variety of morphologies that mix the features of the original training classes.

\vspace{-4mm}

\vspace{-6mm}

\subsection{\label{sec:gspy_glitch_examples}Extracted Gravity Spy Glitches}
\vspace{-6mm}

\begin{figure}[H]
\centering
\includegraphics[width=0.95\textwidth]{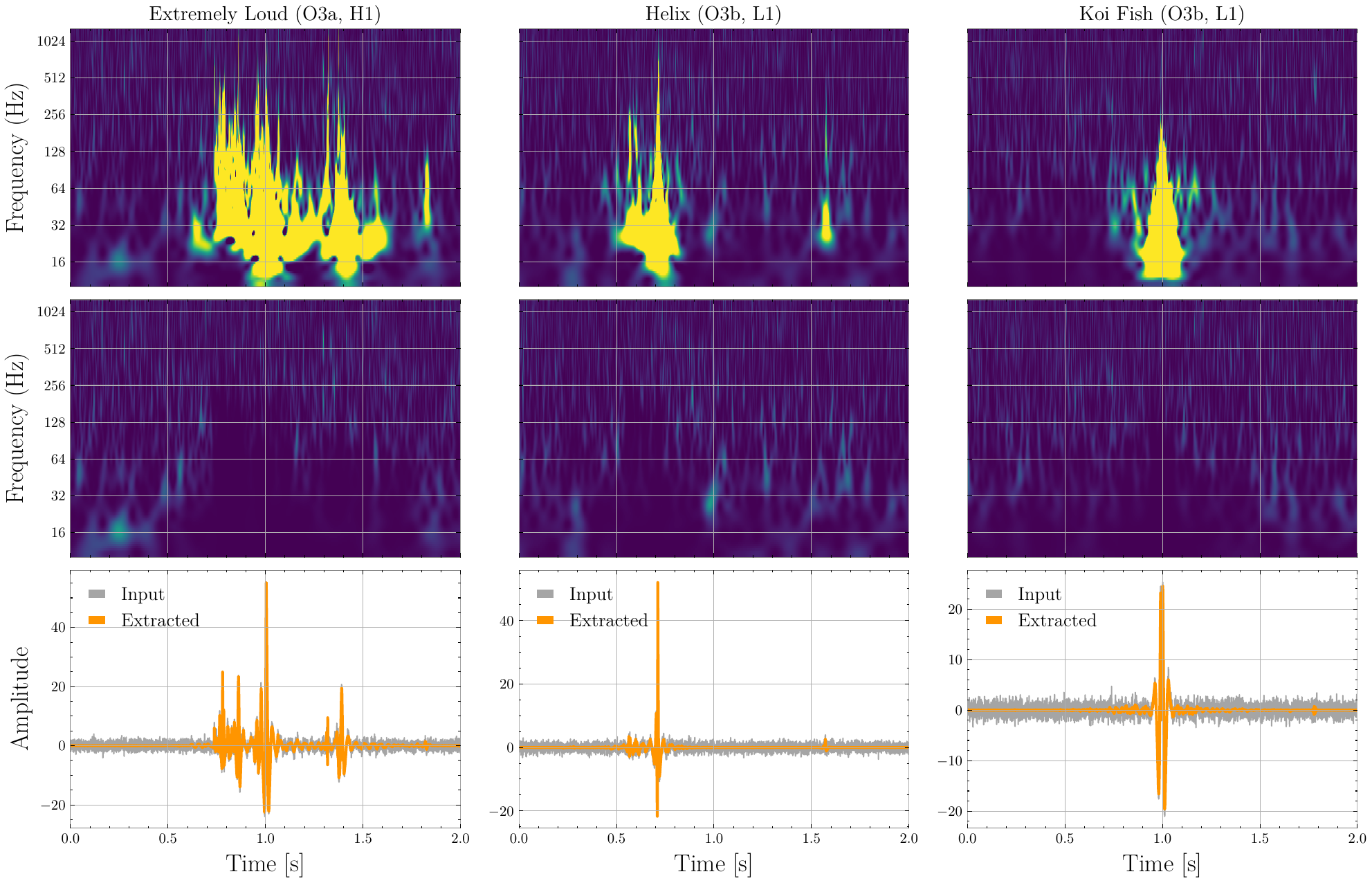}
\vspace{-4mm}
\caption{}
\label{fig:gspy_examples_2}
\end{figure}

\begin{figure}[H]
\centering
\includegraphics[width=0.95\textwidth]{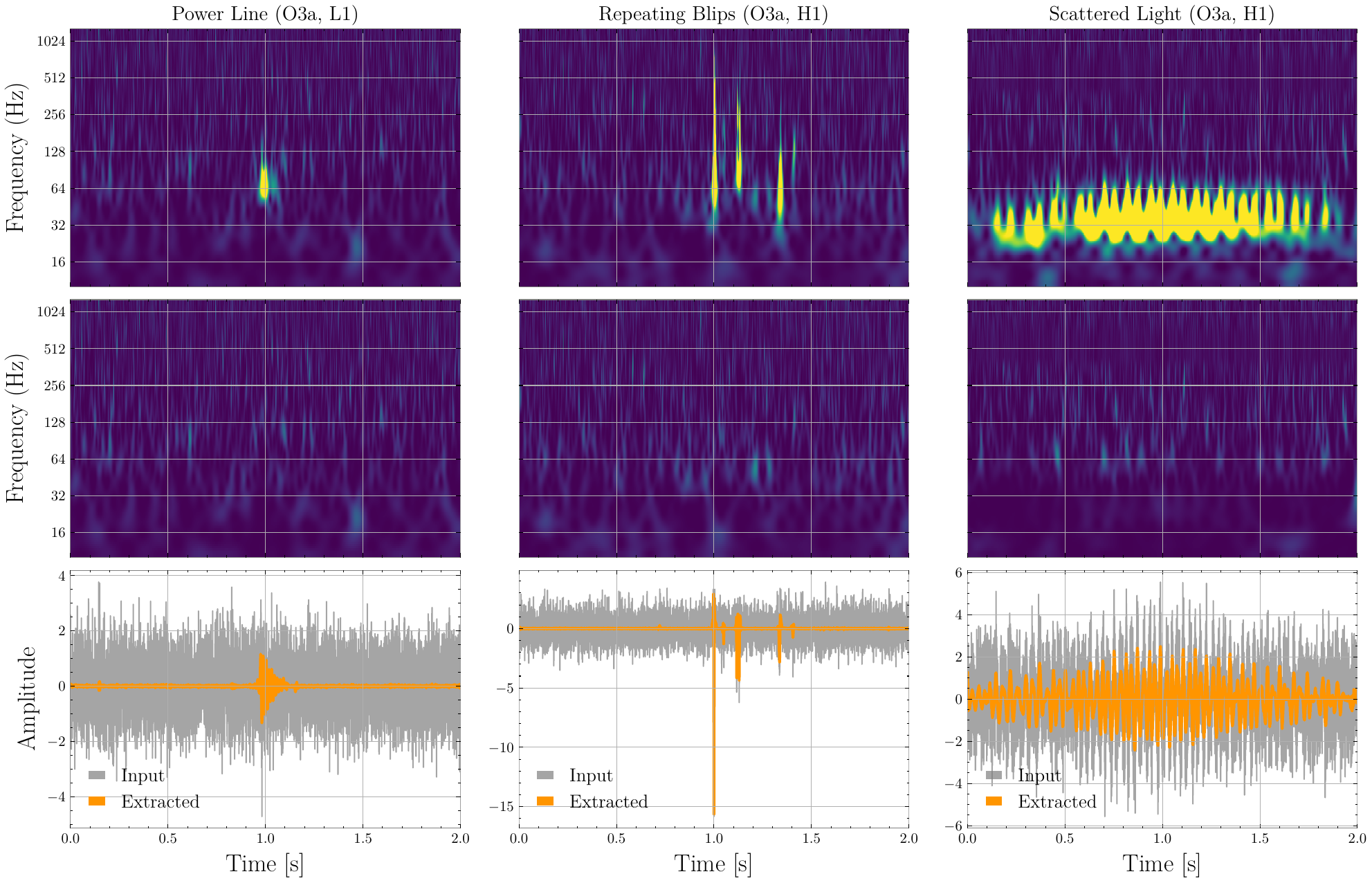}
\vspace{-4mm}
\caption{}
\label{fig:gspy_examples_4}
\end{figure}

\begin{figure}[H]
\centering
\includegraphics[width=0.95\textwidth]{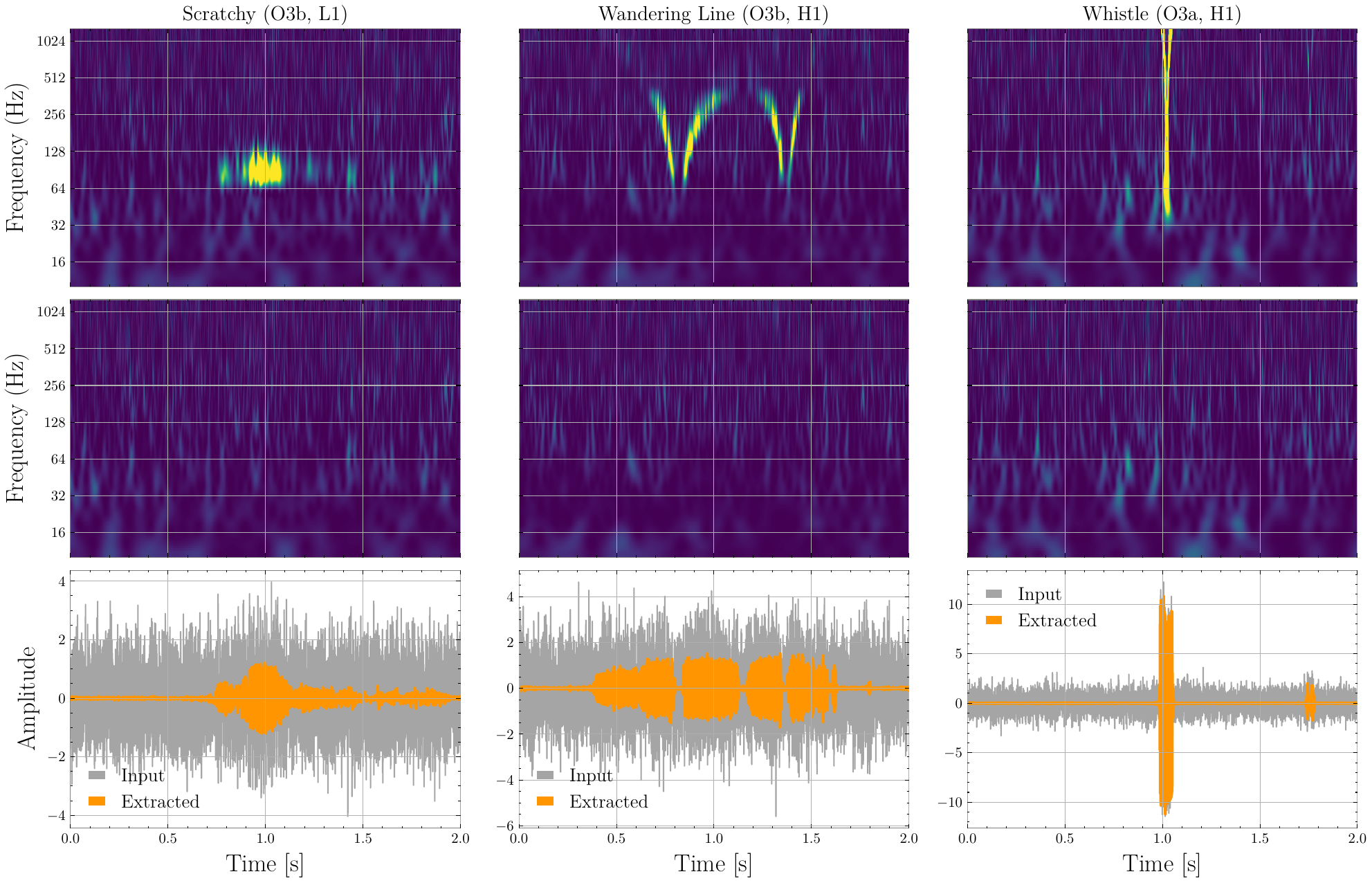}
\vspace{-4mm}
\caption{}
\label{fig:gspy_examples_5}
\end{figure}

\noindent FIG. \ref{fig:gspy_examples_2}, \ref{fig:gspy_examples_4} and \ref{fig:gspy_examples_5} show \textit{DeepExtractor} reconstructions for glitch classes from the \textit{Gravity Spy} dataset. Each figure shows reconstructions for three glitch classes which are named above each column. The top row shows a Q-scan of the input to the network. The middle row shows a Q-scan after removing \textit{DeepExtractor's} reconstruction. The bottom plots show the time series of the input and the reconstruction. The maximum color limit in the Q-scans is set to 25, similarly to the Q-scans shown in FIG. \ref{fig:gspy_transfer_learning}.

\vspace{-3mm}

\vspace{-4mm}

\vspace{-4mm}

\subsection{Reconstructing O3 GWs with glitches\label{sec:gw_glitch_appendix}}

\begin{figure}[H]
\centering

\begin{subfigure}{\linewidth}
    \centering
    \includegraphics[width=\linewidth]{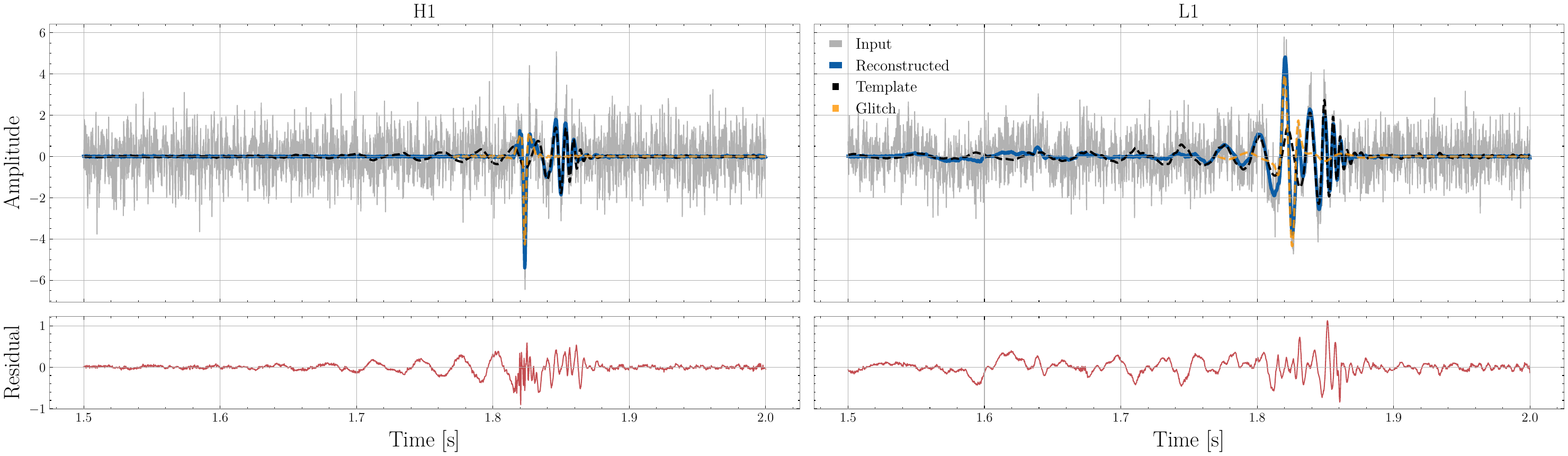}
    \caption{\textit{GW190521\_074359}}
    \label{fig:gw190521_glitch}
\end{subfigure}

\vspace{1em}

\begin{subfigure}{\linewidth}
    \centering
    \includegraphics[width=\linewidth]{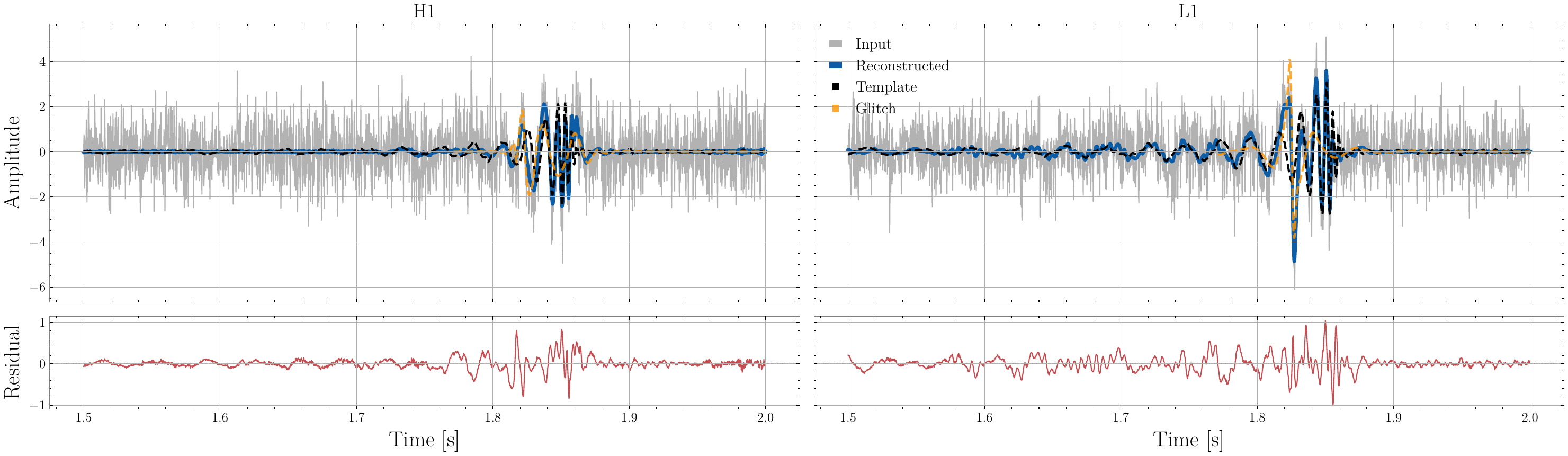}
    \caption{\textit{GW200129\_065458}}
    \label{fig:gw200129_glitch}
\end{subfigure}

\vspace{1em}

\begin{subfigure}{\linewidth}
    \centering
    \includegraphics[width=\linewidth]{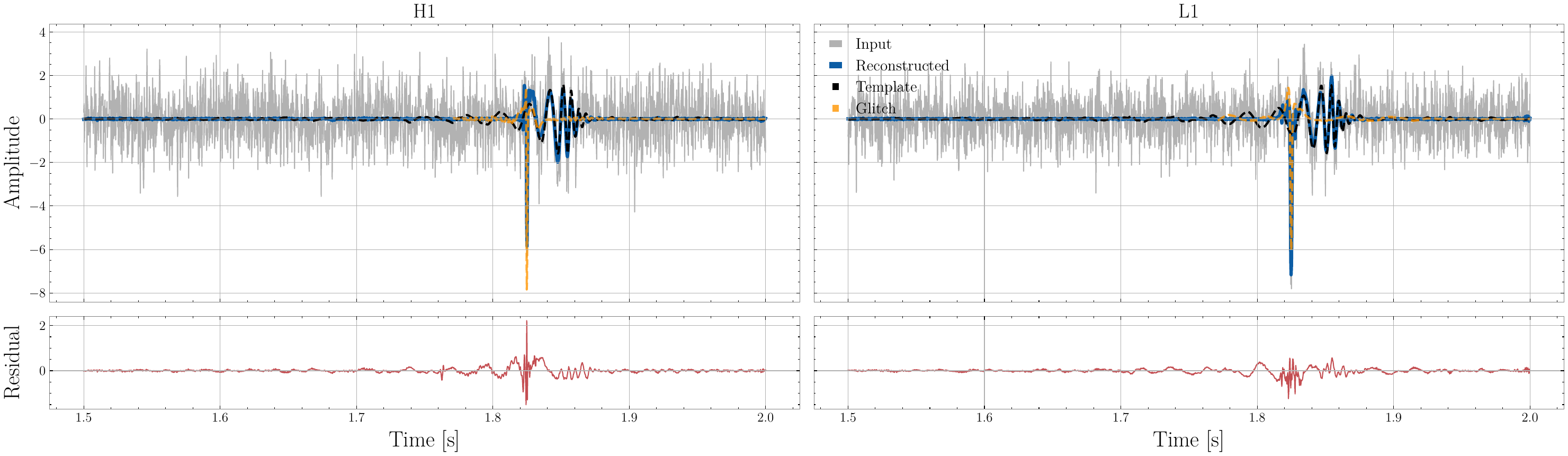}
    \caption{\textit{GW200224\_222234}}
    \label{fig:gw200224_glitch}
\end{subfigure}

\caption{\textit{DeepExtractor} reconstructions of gravitational-wave signals from the same three O3 events shown in Section \ref{sec:gw_results}, now with blip glitches injected near the merger in each detector. Glitches are injected with SNRs comparable to the matched-filter SNR of the corresponding maximum likelihood waveform. Each panel displays the model reconstruction overlaid with the aligned template and injected glitch. The residual plot compares the reconstruction to the linear combination of the signal and glitch.}
\label{fig:GW_reco_glitch_all}
\end{figure}

FIG. \ref{fig:GW_reco_glitch_all} shows \textit{DeepExtractor} reconstructions for the same events shown in Section \ref{sec:gw_results} but with blip glitches injected close to the merger using \textit{gengli}. It is observed that when glitches are coincident with GWs, that \textit{DeepExtractor} will reconstruct both sources of excess power without distinguishing between them.

\newpage

\bibliography{apssamp}

\end{document}